\documentclass[a4paper,fleqn]{cas-dc}

\usepackage[authoryear,longnamesfirst]{natbib}
\usepackage{todonotes}
\usepackage{caption}
\usepackage{subcaption}
\usepackage{xcolor}

\def\tsc#1{\csdef{#1}{\textsc{\lowercase{#1}}\xspace}}
\tsc{WGM}
\tsc{QE}
\tsc{EP}
\tsc{PMS}
\tsc{BEC}
\tsc{DE}

\begin{document}
\let\WriteBookmarks\relax
\def\floatpagepagefraction{1}
\def\textpagefraction{.001}

\shorttitle{Morphological Study of Granular-Granular Impact Craters through Time-of-Flight Cameras}
\shortauthors{F. Corrales et~al.}

\title [mode = title]{Morphological Study of Granular-Granular Impact Craters through Time-of-Flight Cameras: from Concept to Automation in Python}

\author[1]{F. Corrales-Machín}[orcid = 0000-0002-6980-1813]
\cormark[1] 
\ead{frankcm.work@gmail.com}  
\credit{Software, Writing of the manuscript}

\author[2]{G. Viera-López}[orcid = 0000-0002-9661-5709]
\credit{Software, Writing of the manuscript}

\author[3]{R. Bartali}[orcid = 0000-0002-2321-9334]
\credit{Discussion, Writing of the manuscript}

\author[1]{Y. Nahmad-Molinari}[orcid = 0000-0002-3477-2282]
\credit{Conceptualization of this study, Methodology, Writing of the manuscript}

\address[1]{Universidad Autónoma de San Luis Potosí, Instituto de Física, Av. Parque Chapultepec 1570, San Luis Potosí 78295, México}
\address[2]{Gran Sasso Science Institute, Viale Francesco Crispi, 7, L'Aquila 67100, Italy}
\address[3]{Universidad Autónoma de San Luis Potosí, Facultad de Ciencias, Av. Parque Chapultepec 1570, San Luis Potosí, 78295, México} 

\begin{abstract}
  Laboratory made granular-granular impact craters have been used as model analogues of planetary impact craters. These kind of craters have been observed and studied using profilometry techniques that allow to retrieve important morphologic features from the impacted surface. In this work, we propose to use a Time-of-Flight camera (Microsoft Kinect One) for the acquisition of depth data. We show comparisons between the typically used technique and the analysis derived from the Time-of-Flight data. We also release \emph{craterslab}, a Python library developed to automate most of the tasks from the process of studying impact craters produced by granular projectiles hitting on the surface of granular targets. The library is able to acquire, identify, and measure morphological features of impacted surfaces through the reconstruction of 3D topographic maps. Our results show that using a Time-of-Flight camera and automating the data processing with a software library for the systematic study of impact craters can produce very accurate results while reducing the time spent on different stages of the process.
\end{abstract}

\begin{highlights}
\item Introduces a robust technique based on ToF sensors for studying experimental man-made craters morphology.
\item Compares the results obtained with ToF sensors against the established technique.
\item Proposes a software library for the automation of impact craters morphology and morphometric measurements.
\end{highlights}


\maketitle
\section{Introduction}
\label{intro}

Determining unknown distances to objects or their spatial dimensions by measuring the angles they form from known points of observation is an ancient technique known as triangulation, which is still used in modern instruments. In the sixth century BC, Thales of Miletus measured the height of pyramids by comparing the ratio of their shadow length to height with his own shadow length to height ratio at the same time, using the Thales theorem of corresponding proportions. Shortly after, Eratosthenes measured the radius of the Earth, while Aristarchus of Samos calculated the sizes of the Sun and Moon, as well as their distances from the Earth in Earth radii, based on the same geometric principles. This led to the development of a heliocentric model of our solar system, utilizing a simple yet powerful set of geometric theorems, thousands of years ago \citep{james1953historia}.
	
After Galileo invented the astronomical telescope in $1609$ and discovered craters on the Moon's surface, various hypotheses were proposed regarding the origin of these geological structures, including coraline reefs, volcanic activity, and the later rejected idea of impact origin proposed by Hooke. It was not until the twentieth century that the impact origin theory was revived by Grove Gilbert for explaining lunar craters, and found to align well with Laplace's protoplanetary cloud theory of solar system formation \citep{gilbert1979moon}. 
	
The importance of meteorite impacts for Earth and life on Earth evolution was recognized in $1980$, when the Chicxulub crater in the Yucatán peninsula was recognized as the scar of a colossal impact that caused the mass extinction event at the Cretaceous - Paleogene (K-Pg) boundary $65$\,Ma ago \citep{alvarez1980extraterrestrial,alvarez1995emplacement}. Initially, projected shadow length was used to determine the depth of craters and the height of their rims in early studies of lunar geophysical features \citep{chappelow2013simple}. Subsequently, satellite radar altimetry using real-time of flight techniques \citep{davis1992satellite} was employed to explore topographic features and create elevation maps. Eventually, phase-change Time-of-Flight techniques, such as LiDAR, were introduced for atmospheric, terrestrial, and planetary science prospecting.
	
Currently, there exists a well-established understanding of the processes involved in impact crater formation, which has been derived from geophysical exploration of terrestrial impact craters, computer simulations, and hypervelocity experiments. These processes can be categorized into three main stages: contact and compression, excavation of a transient crater, and modification through avalanching and deposition of debris \citep{melosh1989impact,osinski2013impact}. However, due to their rare occurrence and the immense energy involved, impacts that form planetary craters are infrequent events that are difficult to observe directly. Consequently, it is challenging to gather experimental or observational evidence to directly compare and validate the theoretical understanding of impact crater formation.
	
Again, using proportionality laws or scaled systems, the Scottish geologist and geographer Henry Cadell played a pivotal role in advancing the field of analog model studies through sandbox experiments. His work focused on investigating the formation of thrust and fold systems in the Scottish Highlands. Subsequently, scaled analogue modeling has become a commonly employed technique for studying the geometric, kinematic, and dynamic evolution of various geological structures. This powerful tool allows for a comprehensive understanding of the geometric and kinematic development of extensional, inverted fault systems, as well as strike-slip fault systems. 
	
The remarkable resemblance between the scaled models and the natural geological examples described in the literature highlights the effectiveness of this method in accurately replicating real-world geological structures \citep{mcclay1996recent}. However, this technique has been, just very recently, incorporated for understanding Impact craters as geologic processes \citep{bartali2015low} by considering equal adimensional numbers (v.g. Reynolds and comminution numbers), regardless of the fact that man made laboratory craters and observed planetary craters are produced by events differing in six or more orders of magnitude \citep{gonzalez2014penetration,pacheco2011impact}.	
		
In order to investigate the influence of impact collision energy on the final shape of craters, various techniques have been employed for the characterization of morphological features of craters such as laser profilometry or direct measurements \citep{de2007shape}. These techniques provide valuable insights into the characteristics and behavior of impact craters, aiding in the understanding of the relationship between the energy involved in the event and the resulting crater morphology and sedimentologic features.
		
In 2010, Microsoft released a structured light-based range detection camera, Kinect, which provides depth images (RGB-D) along with an RGB color camera. Although the Kinect sensor was originally intended for natural user interaction in body-based video games, the release of its source code by Microsoft has led to the development of numerous applications in robotics \citep{el2012study}, 3D reconstruction \citep{keller2013real, newcombe2011kinectfusion, niessner2013real}, medicine \citep{mousavi2014review}, augmented reality and interaction \citep{vera2011augmented}, geophysics \citep{rincon2022monitoring, tortini2014innovative}, among others.

In 2013, Microsoft announced an update to Kinect based on the Time-of-Flight (ToF) principle. This new version includes additional improvements compared to its predecessor. 	
	
In the study of craters, the Kinect system has been employed to automatically measure grain size distribution across a range from pebbles to blocks in outcrops within the Joya Honda crater in Mexico \citep{chavez2014using}. However, the increasing utilization and affordability of LiDAR and Time-of-Flight instruments for rapid surface topography measurement have prompted us to develop a versatile methodology specifically designed for acquiring and processing topographic data in the study of impact crater formation. 
	
As part of this work, we release a Python library we develop to automate our methodology and determine the morphological characteristics of excavated craters in laboratory settings. We expect that both our library and our approach on using Time-of-Flight cameras may enable novel studies on granular-granular impact craters serving as model analogues for observed planetary craters.

\section{Mapping of surfaces}
\label{''mapping''}
	
Three-dimensional measurement and reconstruction of surfaces is a significant topic in various fields of research, with diverse applications such as range scanning \citep{zhang2009high}, industrial inspection of manufactured parts \citep{graebling2002optical}, reverse engineering (digitization of complex, free-form surfaces) \citep{lu2015multi,carbone2001combination}, object recognition and 3D mapping \citep{stein1992structural,rogers2011simultaneous}. Currently, several techniques are implemented for these measurements, benefiting from significant technological advancements that enable high resolutions and software with multiple domain-specific features. However, access to these software often comes with a high financial cost.
	
In the context of mapping  granular-type impact craters, the scientific community primarily relies on profilometry as the preferred technique for obtaining morphological characteristics. However, the idea of implementing depth measurement techniques based on range sensors, such as LiDAR, in this field of research is highly appealing. In this section, we will explain the operating principle of both techniques and their general limitations, with a deeper focus on their application for the study of craters morphology.
	
\subsection{Profilometry-based Methods}	
	
With the current technological advances in acquiring three-dimensional surface maps, different profilometry techniques have been refined to obtain more reliable results in shorter time \citep{van2016real, salvi2010state, su2001fourier}. Despite these advancements, most of these techniques are challenging to implement and have limitations such as complex image analysis. 
	
As mentioned earlier, laser profilometry is commonly used to obtain morphological characteristics of craters. This method is based on the principle of triangulation, where a laser projects a beam of light onto the surface of interest, and a sensor records the position and angle of the reflected beam. With this information, the distance between the sensor and the surface can be calculated, allowing for the reconstruction of a three-dimensional profile. 
	
In addition to laser profilometry, another technique used for measuring depths on crater surfaces is structured light profilometry \citep{geng2011structured}. In this method, a pattern of structured light, such as stripes or lines, is projected onto the surface, and an image of the illuminated surface is captured. Analyzing the deformations of the light pattern in the image allows for the calculation of local depths of the surface. Structured light profilometry is based on the principle of interferometry, where variations in the surface shape cause changes in the phase and intensity of the reflected light. These changes are captured by a camera and processed to obtain a depth map of the crater's surface.
	
While laser profilometry and structured light profilometry are widely used techniques for obtaining data for the morphological characterization of granular impact surfaces, they also have certain limitations that are important to consider. The following are some of these limitations:
	
Both laser profilometry and structured light profilometry methods have limitations in resolution due to factors such as sensor-to-surface distance, pixel size, and laser precision. These limitations can impede capturing fine surface details, especially in areas with small features. Additionally, accurately measuring transparent or translucent surfaces can be challenging as light may pass through or be absorbed instead of being reflected, resulting in inconsistent measurements. Reflective surfaces can also pose difficulties, as intense reflections can interfere with measurements and generate inaccurate data. Shadows and obstructed areas on the surface can hinder data capture by causing variations in reflected light intensity or blocking the light pattern projection. Furthermore, measurements obtained through these methods are susceptible to noise and artifacts, which can introduce errors or distortions in the data. These artifacts can arise from fluctuations in light intensity, environmental interference, or device calibration issues. Finally, data acquisition time can be a limitation, particularly when high resolution or sampling large areas efficiently is required, impacting situations that demand fast response times.
	
In summary, laser profilometry and structured light profilometry are valuable techniques for measuring depths and obtaining three-dimensional surface information. While they have seen improvements in recent years, they still have limitations in terms of implementation complexity and specific challenges related to image analysis. These techniques, nevertheless, offer valuable insights into the morphology of granular impact craters and contribute to the understanding of physical phenomena.
	
\subsection{Methods based on LiDAR Sensors}	
\label{sub:lidar}	
	
In the last decade, new affordable range detection devices have been developed. Light Detection and Ranging (LiDAR), since the 1960s with the advent of lasers, has emerged as a pioneer in this field, empowering multiple applications \citep{dong2017lidar, pittman2013lidar}. LiDAR technology is based on the Time-of-Flight principle. It measures the time it takes for light emitted by a device to travel to the surface of an object and return to the sensor of the unit. The precise measurement of the time it takes for light to travel and return to the sensor array of a measuring device is determined by the switching velocity of the sensor's microelectronics. Time-of-Flight cameras employ a continuous wave intensity modulation approach, where the surface of interest is illuminated with near-infrared intensity-modulated periodic light. Considering the finite speed of light ($c$) and the distance between the camera and the surface (assuming the sensor and illumination are in the same location), an optical signal experiences a temporal shift $\phi[d]$, which corresponds to a phase shift in the periodic signal. The phase shift is calculated by considering the charge accumulated in the sensor due to the reflected light when the synchronous shutter turns off the light sampling. By transforming the temporal shift into the sensor-object distance, we obtain the equation $d = 4c\phi\pi$. It is important to note that intermittent illumination of the scene at several gigahertz and rapid switching speeds are crucial for achieving high depth resolution.
	
Among the various LiDAR devices based on the Time-of-Flight principle, the second generation of Microsoft Kinect (KinectToF) stands out. It offers several improvements over its predecessor, which utilizes structured light (SL) method for depth information acquisition. In the first generation of Kinect, the structured light method involves projecting a sequence of known patterns onto an object, which deform based on the object's shape. The deformed patterns are then captured by a camera, and by analyzing the distortion using triangulation, depth information is derived.
	
Both SL and ToF principles for range detection are susceptible to various sources of error. Several studies have compared these methods and explored different calibration techniques for the Kinect camera. These studies include \citep{sarbolandi2015kinect, wasenmuller2017comparison, pagliari2015calibration, yang2015evaluating, lachat2015first, essmaeel2012temporal, zhang2000flexible}. Considering the benefits and limitations of the two different Kinect principles of operation, it has been determined that the second generation, utilizing ToF technology, is superior \citep{kadambi20143d}. To our knowledge, ToF sensors have not yet been used for the study of morphological signatures of experimental impact craters in laboratory.
	
\section{Materials and Methods}
\label{sect:materiales}

An experimental system was designed to recreate the formation of impact craters by using, for the first time, a KinectToF sensor for the data acquisition. Considering that laser profilometry is the typically used technique for this purpose, we added it to the experimental setup in order to validate the results obtained by our approach. 
		
We constructed a square-based sandbox with dimensions of $45$\,cm per side and $15$\,cm in height as the surface or granular bed in which the crater forms after the impact of a sand lump projectile. Sand grains with a diameter of $d\leq1.0$\,mm were deposited inside the box as the granular medium. The granular bed is loose packed or compacted in order to observe how the morphologies of the craters vary for different impact energies.
	
Impacts were carried out by releasing a granular projectile from heights ranging from $0.1$\,m to $20$\,m, respectively. The granular projectiles were composed of $250$\,g of the same granular material as the impact surface, using $50$\,ml of water, and $5.0$\,g of hydraulic Portland cement as an adhesive. The mixture was compacted into a spherical mold and left to dry at room temperature, forming weakly consolidated granular spheres with a diameter of $7.0$\,cm and packing fractions ranging from $\eta=0.40$ to $\eta=0.62$.
	
For retrieving depth maps from the granular surface, we attached a Microsoft Kinect One to a mobile system placed over the sandbox, allowing the sensor to move along one horizontal (the $Y$ axis) above the sand free surface during experiments. Depth data was acquired at a height of $102.7$\,cm, perpendicular and stationary to the impact surface. Two depth maps are acquired using Kinect, the first one containing all projectile fragments that may be present on the surface, and for the second one, the interfering projectile fragments are removed to facilitate the morphological analysis of the impacted surface. Once the depth data is acquired using the Kinect sensor, a custom software was used in order to process the data and retrieve valuable information from the surface.
		
The laser profilometry technique is performed as well in order to compare and discuss the results of both methods. 	It is conducted without a sensor for automated data acquisition. Instead, a laser beam is used to project five lines onto the granular bed at a $45$-degree angle. Scanning is performed at different points on the surface, and images are captured for each position. Subsequently, these images are processed using \emph{ImageJ} software \citep{bartali2013role}, employing the principle of triangulation to obtain depth and diameter measurements of the crater under study. The procedure of using laser profilometry to obtain morphological characteristics is well-known and established in the field.
	
Next, we will address some definitions related to morphological observables from impact craters.

\section{Main Crater Observables}
\label{sect:observables}	

\begin{figure}[ht]
\centering
\begin{subfigure}[ht]{0.23\textwidth}
\centering
\includegraphics[width=\textwidth]{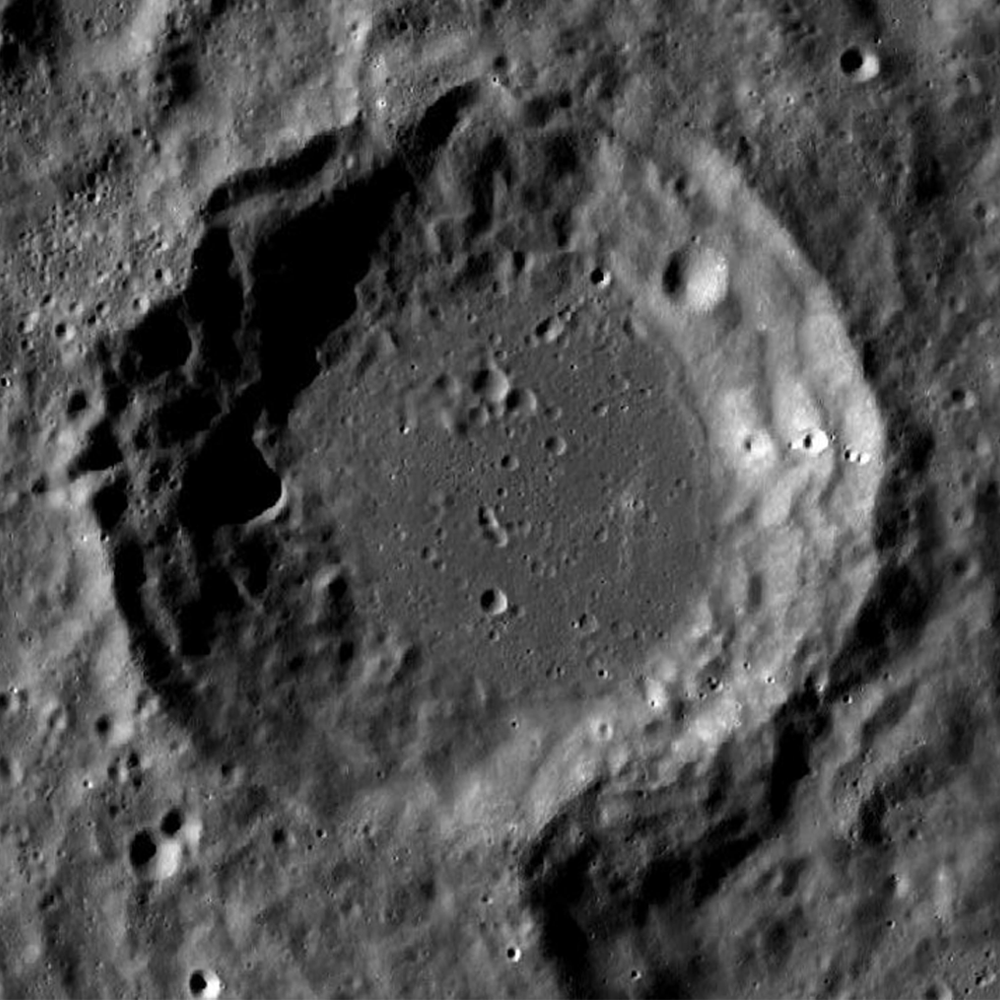}
\caption{}
\label{fig:1a}
\end{subfigure}
\begin{subfigure}[ht]{0.23\textwidth}
\centering
\includegraphics[width=\textwidth]{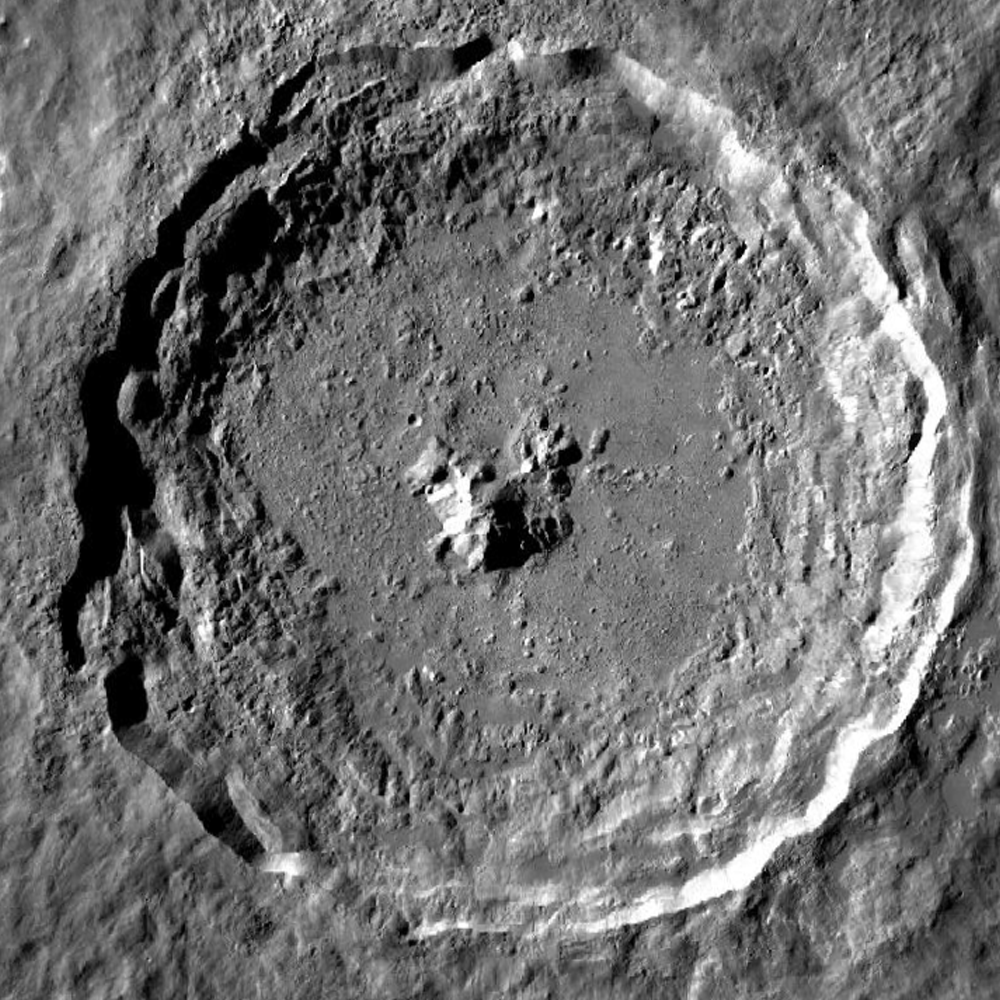}
\caption{}
\label{fig:1b}
\end{subfigure}
\caption{Crater classification. 
(a) Simple crater Steinheil  
(b) Complex crater Tycho. Images taken from \citep{lroc_steinheil,lroc_tycho}}
\label{fig:Figure1}
\end{figure}

Craters can be classified into two groups: simple and complex craters. Examples from both types can be inspected in Figure \ref{fig:1a} and Figure \ref{fig:1b} respectively.

Simple craters are bowl-shaped depressions with raised rims and approximately parabolic interior profiles. A straightforward structure that presents a circular or elliptical rim with the rim-to-floor depths of a large sample of simple craters on the moon are roughly $1:5$ of the rim-to-rim diameter.
	
Complex craters possess a variety of features and a more complicated structure than simple craters. They often exhibit a central structure in their interior (central peak or dome) which may protrude above the crater rim.  Images from Lunar Reconnaissance Orbiter Camera (LROC) shows craters with single or multiple central peaks, concentric rims and flat inner floors. The depths of complex craters increase with increasing diameter, but they increase much more slowly than the depths of simple craters \citep{melosh1999impact}.

In both cases, we will establish the original ground surface as the zero reference for heights and depths \citep{osinski2013impact}. From this reference point, we will consider the positive $Z$ axis as an increase in height above the surface, and the negative axis as a decrease in surface level.

\begin{figure}[ht]
\centering
\begin{subfigure}[ht]{0.45\textwidth}
\centering
\includegraphics[width=\textwidth]{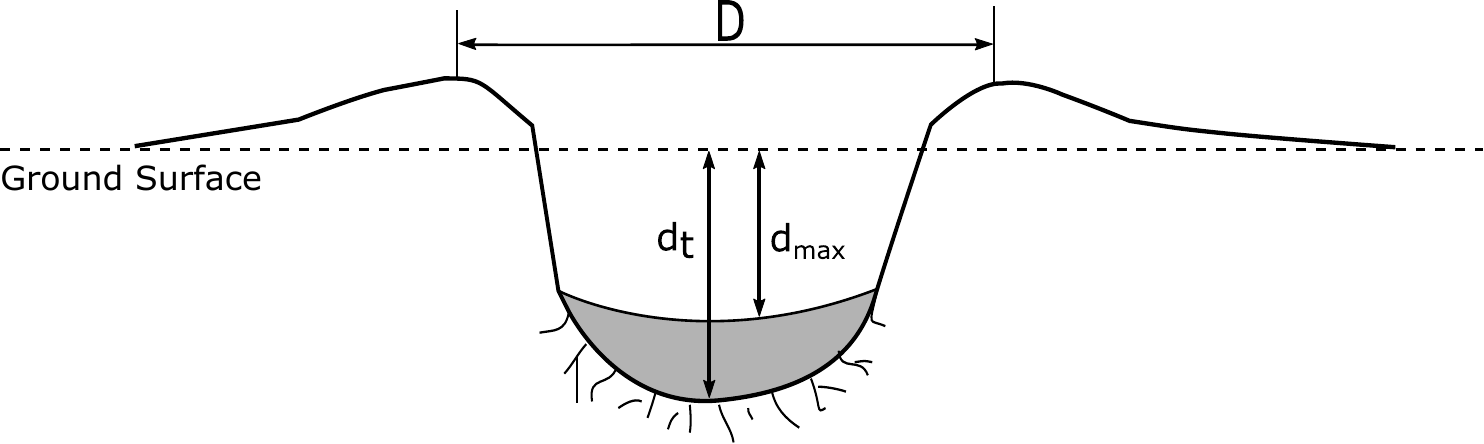}
\caption{}
\label{fig:2a}
\end{subfigure}
\begin{subfigure}[ht]{0.45\textwidth}
\centering
\includegraphics[width=\textwidth]{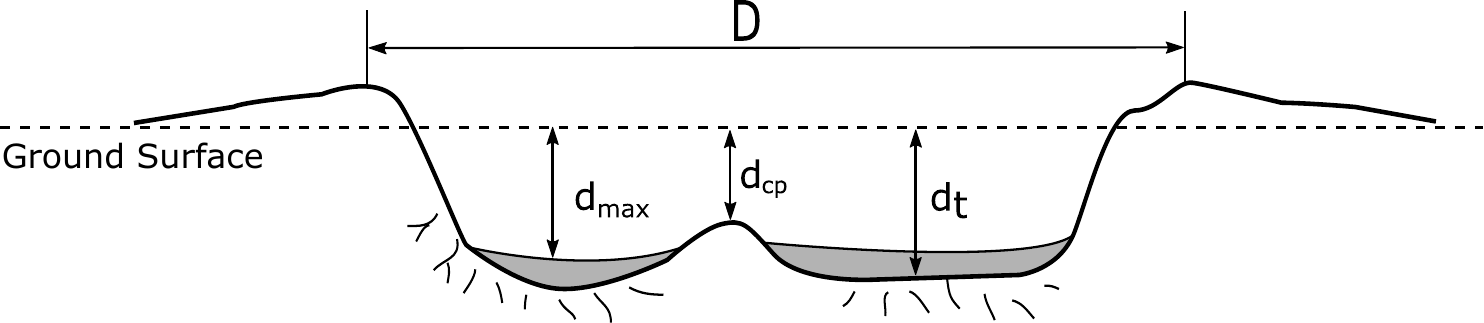}
\caption{}
\label{fig:2b}
\end{subfigure}
\caption{Definition of some craters observables. 
  (a) For Simple Craters. 
  (b) For Complex Craters.}
\label{fig:Figure2}
\end{figure}

For both morphologies depicted in Figure \ref{fig:Figure2} there may be deposits of granular material in the interior of the crater, which are remnants of the impacting granular projectile. Therefore, the maximum observable depth $d_{max}$ may be smaller than the actual crater depth $d_t$. Both measurements of depth are typically below the original ground surface.

The height of the central peak $H_{cp}$ can be defined as the difference between its maximum depth value $d_{cp}$ and the maximum observable depth $d_{max}$. 

From Figure \ref{fig:Figure1} we can notice that craters generally have an elliptical geometry, where a circular approximation of their surface is a special case of an ellipse. To determine the type of geometric approximation that best fits their surface, Equation (\ref*{eqn:elipse}) can be fitted on samples from the distribution of maximum height values around the crater rim $h_{rim}$. 

Once this ellipse is obtained, the values of $a$ and $b$ are fixed, which represent the major and minor radii of the ellipse, respectively. In the case of a circular fit, these values will be equal. Additionally, $x_c$ and $y_c$ represent the $x$ and $y$ coordinates of the center of the ellipse. Finally, Equation \ref{eqn:elipse} fitted over the rim provides the variables $\theta$ and $\varepsilon = \dfrac{\sqrt{a^2-b^2}}{a}$, which are the rotation angle and eccentricity of the ellipse that represents how circular or elliptical the surface of the crater is in function of it diameters.

\begin{equation} \tag{1} \label{eqn:elipse}
\begin{aligned}
\frac{[(x - x_c)\cos\theta + (y-y_c)\sin\theta]^2}{a^2} + \\ \frac{[(x - x_c)\sin\theta - (y-y_c)\cos\theta]^2}{b^2} = 1
\end{aligned}
\end{equation}
	
The diameter $D = 2a$ of both simple and complex craters is defined as the distance equal to the major axis of the best fitting ellipse to the rim. Having an elliptical approximation for a crater simplifies the computation of several observables. For example, the crater diameter $D$ can be conveniently expressed as $D = 2a$. Furthermore, by transforming Equation (\ref{eqn:elipse}) into an inequality, it is possible to quickly determine whether an arbitrary coordinate $x, y$ corresponds to the interior of the crater or not. This can be used to speed up some costly computations, such as finding the maximum observable depth $d_{max}$ or computing the crater concavity's volume $V_{in}$.

The concavity's volume $V_{in}$ is the volume contained inside the crater limited to the average value of $h_{rim}$. This volume is equivalent to the amount of water that can be contained within the crater's concavity if it could be filled up to the average rim's height without being spilled out, considering the rim's height to be uniform all around the fitting ellipse.
	
The excavated volume $V_{ex}$ is the volume of the hollow under the surface reference ground level within the crater rims.
This excavated volume only accounts for the amount of material of the target that has been removed or compressed, but not substituted by the projectile material. As the impact energy increases, larger is the excavated volume, and less material from the projectile remains within the crater.

For complex craters with uplifted central structures, such as central peaks or domes, the crater's depression forms an annular trough. The lowest points of this annular depression delineate a ring-shaped valley, marking the beginning of the uplifted central structure. The volume of the central peak ($V_{cp}$) corresponds to the space enclosed within the inner ring-shaped valley and is determined based on the average depth of this valley.

On the other hand, we define the deposit excess volume $V_{exc}$ as the volume found above the ground surface. This variable is related to the amount of material ejected  and expanded (Reynolds dilatancy \citep{andreotti2013granular}) or displaced by the shock wave during the impact and includes all material deposited on the surface, including that of the crater rim. It is important to note that in the case of complex craters, the formation of central peaks may protrude above the ground surface zero reference, and that fraction of the volume of the central peak $V_{cp}$ is considered in the calculation of $V_{exc}$.

\begin{figure}[ht]
\centering
\begin{subfigure}[ht]{0.23\textwidth}
\centering
\includegraphics[width=\textwidth]{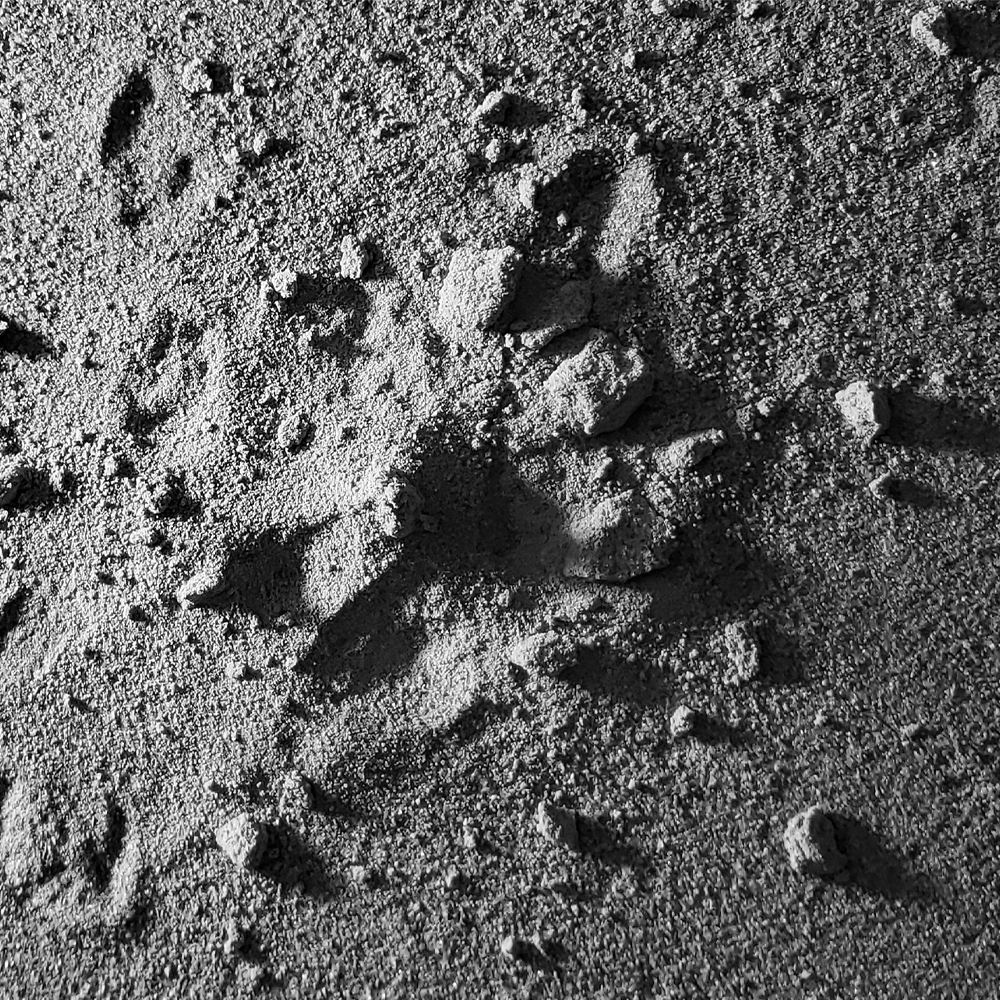}
\caption{}
\label{fig:3a}
\end{subfigure}
\begin{subfigure}[ht]{0.23\textwidth}
\centering
\includegraphics[width=\textwidth]{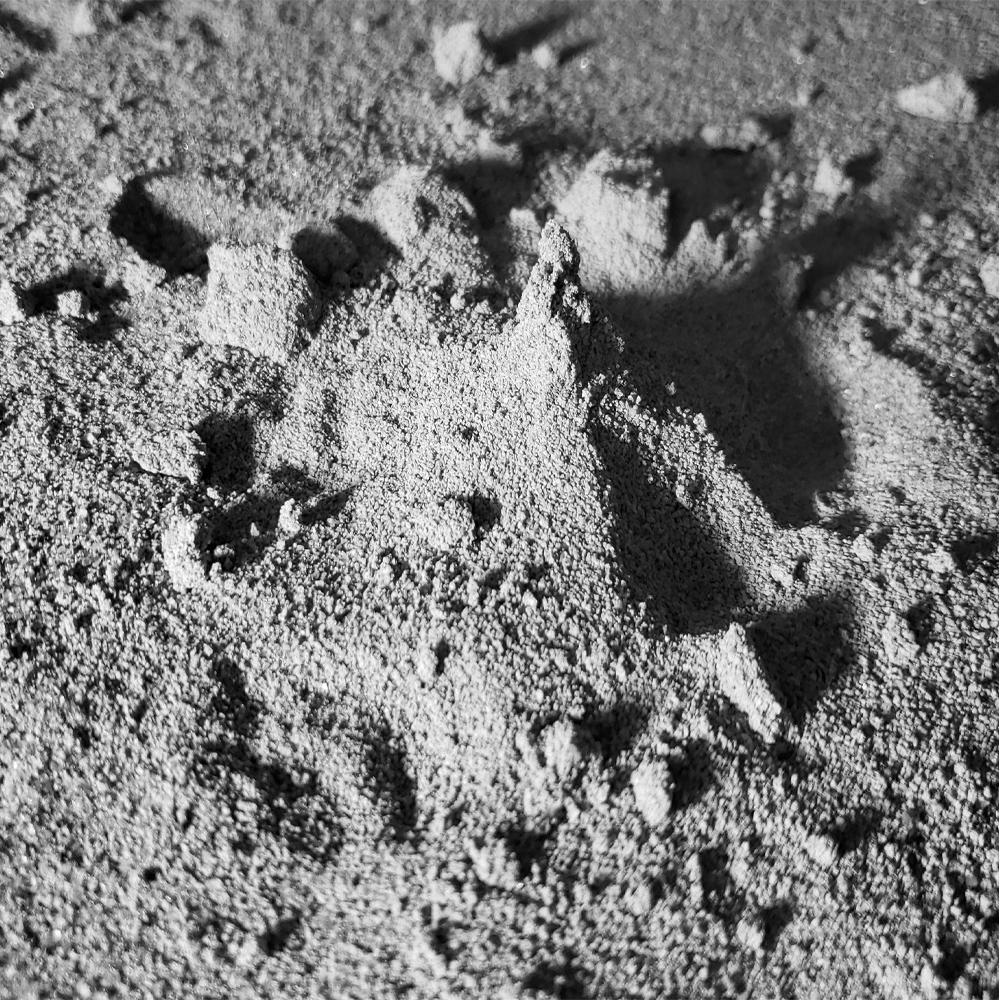}
\caption{}
\label{fig:3b}
\end{subfigure}
\caption{Impacts at very low energies with no penetration of original ground surface.
(a) Perpendicular view respect to the original ground surface 
(b) Oblique view.}
\label{fig:Figure3}
\end{figure}

Now, let's delve into a particular case of morphologies observed for impacts with very low energies, where no penetration occurs in the original and compacted ground surface (See Figure \ref{fig:Figure3}).

These particular morphologies do not meet the definitions of craters, since no depression is formed, and will be considered as sand mounds formed by the remnants of the projectile on the impact surface. These mounds may or may not approximate central peaks. For these cases, only the variables of deposit excess volume $V_{exc}$ and maximum mound height $H_m$ for $z>0$ values are taken into account.

\section{Craterslab Software}
\label{sect:software}
	
The study of planetary geology and impact craters encouraged the development of various software tools that aid in the analysis of planetary craters. These software tools provide valuable insights into the formation and evolution of celestial bodies, helping us to better understand the history and structure of our solar system.

Some of the most recent software tools for analyzing planetary craters include craterstats \citep{craterstats3}, CSFD Tools \citep{riedel2018new}, mvtk \citep{mcdonald2015real} and PyNAPLE \citep{sheward2022pynaple}. These software offer a range of features, from 3D visualization and topographic mapping to data analysis and modeling tools. They are widely used by planetary scientists and researchers to study the morphology and history of craters on various celestial bodies.
		
However,  most of the craters-related software have been crafted with a strong focus on planetary craters. While man-made craters have been shown to be useful models for studying the rare events occurred during impact crater formation, specific software tools are required to help process data from these experiments. To address this need, we have developed \emph{craterslab}, a software library that is able to automate data acquisition from Time-of-Flight sensors and process the data to retrieve the main crater morphologic features. The library is open source and packaged for distribution via pypi \citep{craterslab}.
	
\begin{figure}[ht]
\centering
\begin{subfigure}[ht]{0.23\textwidth}
\centering
\includegraphics[width=\textwidth]{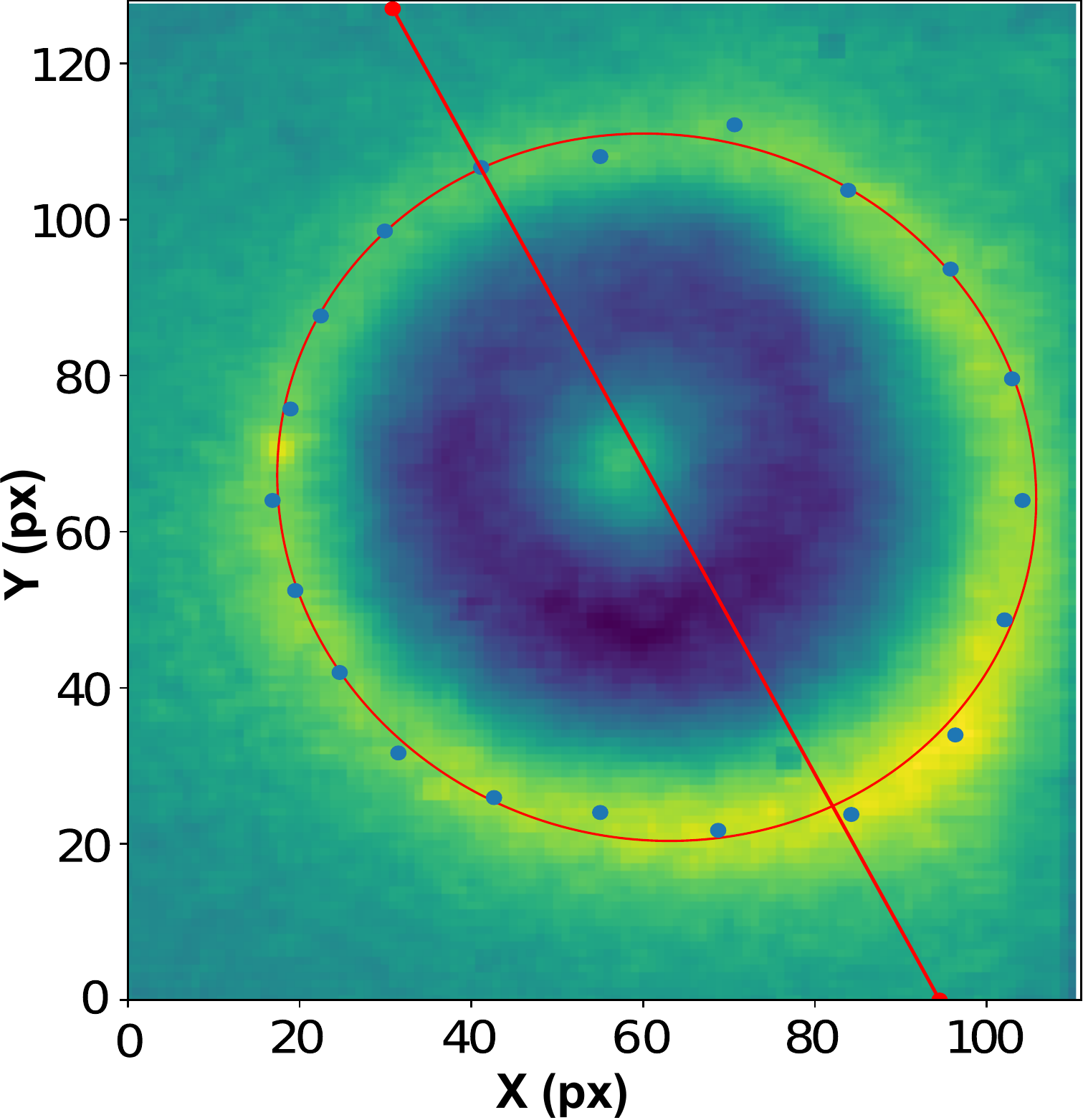}
\caption{}
\label{fig:4a}
\end{subfigure}
\begin{subfigure}[ht]{0.23\textwidth}
\centering
\includegraphics[width=\textwidth]{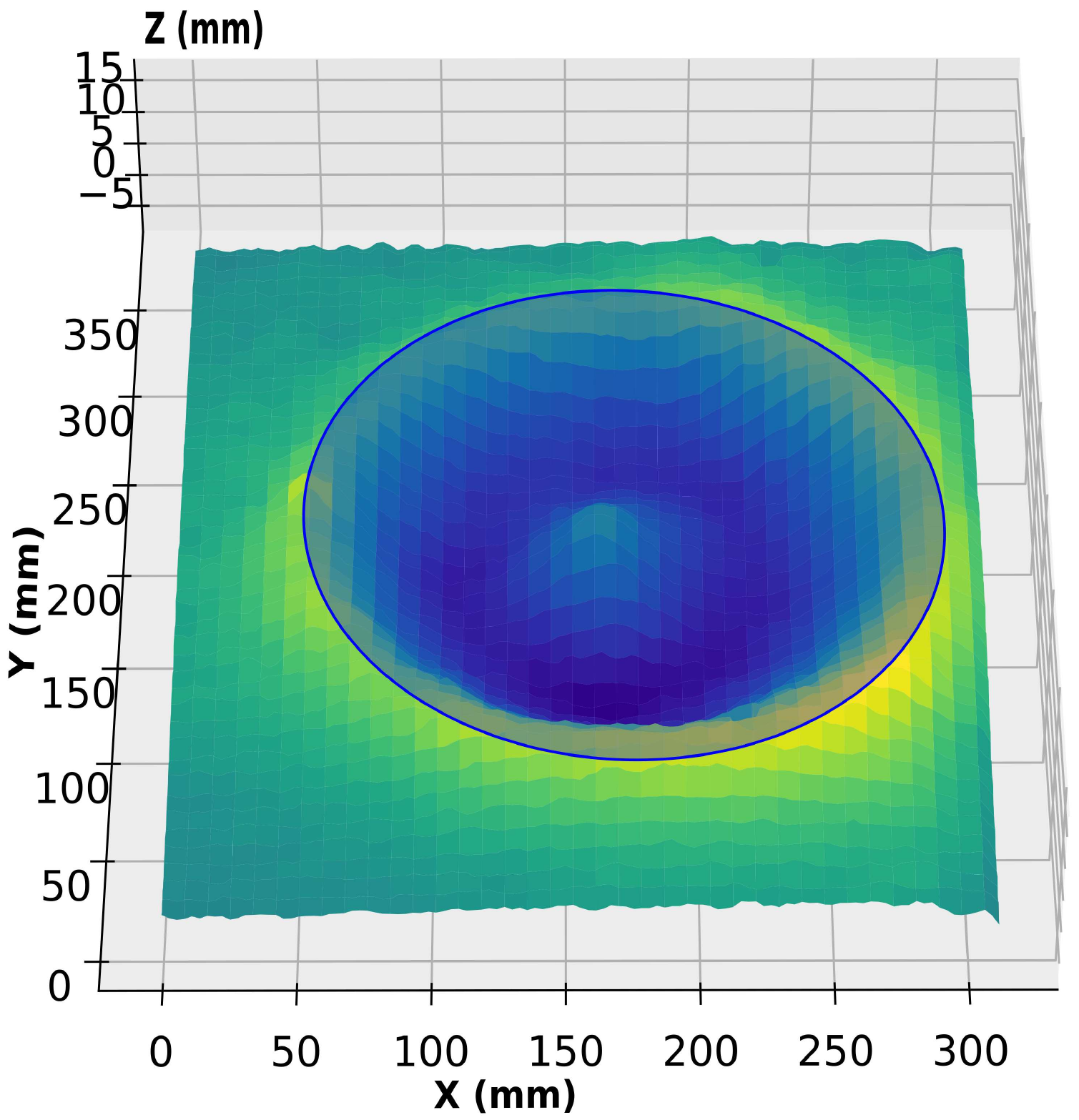}
\caption{}
\label{fig:4b}
\end{subfigure}
\caption{Visualization of the impacted surfaces using \emph{craterslab}.
(a) 2D view of the impact surface in the $X,Y$ plane. The fitted ellipse is observed over the distribution of maximum height values around the crater rim. The diameter, which coincides with the major axis of the fitted ellipse, is also represented.
(b) 3D visualization of the cavity volume, which can be interpreted as the amount of water that can be contained within the crater. This provides a visual interpretation of the numerical value of $V_{in}$.
}
\label{fig:Figure4}
\end{figure}

The software is designed to simplify the data acquisition and analysis process, allowing researchers to focus on the interpretation of results rather than spending time on data processing. It offers a range of features, including automatic data acquisition, real-time data processing, and the ability to visualize and analyze data in a variety of ways. Sample plots produced with \emph{craterslabs} can be seen in Figure \ref{fig:Figure4}.

The workflow of the software can be summarized in: (1) fetching surface mapping data directly from sensors or locally stored files; (2) Classifying the surface based on the observed formation: simple crater, complex crater or sand mound; (3) computing the shape of the crater by fitting an ellipse to the crater rim; (4) determining morphological crater features and (5) visualizing the results. However, the software is built with flexibility, allowing for independent usage of some of its functionalities.

The different crater's observables computed by the software, described in Section \ref{sect:observables}, allows for various analyses of experimental crater morphology. Variables such as diameter and depth can be more accurately correlated with each other. Others, like cavity volume, can now be determined precisely with numerical integration rather than geometric approximations. For example, it is now possible to calculate the volume of the cavity in the craters represented in Figure \ref{fig:5d} and Figure \ref{fig:5e}.

The software is also able to compute the interior slopes of the craters, which allows to determine the incoming direction of the projectile in oblique impacts; the excavated and excess volume, which are related to the amount of material deposited inside the crater, compression and expansion of the terrain, and the ejecta deposited outside the crater after impact.
	
In the following section, we will illustrate the usage of the software by processing the data obtained following the procedure described in Section \ref{sect:materiales}.

\section{Results and Discussion}
\label{sect:discussion}
	
\begin{figure*}[ht]
\centering
\begin{subfigure}[ht]{0.3\textwidth}
\centering
\includegraphics[width=\textwidth]{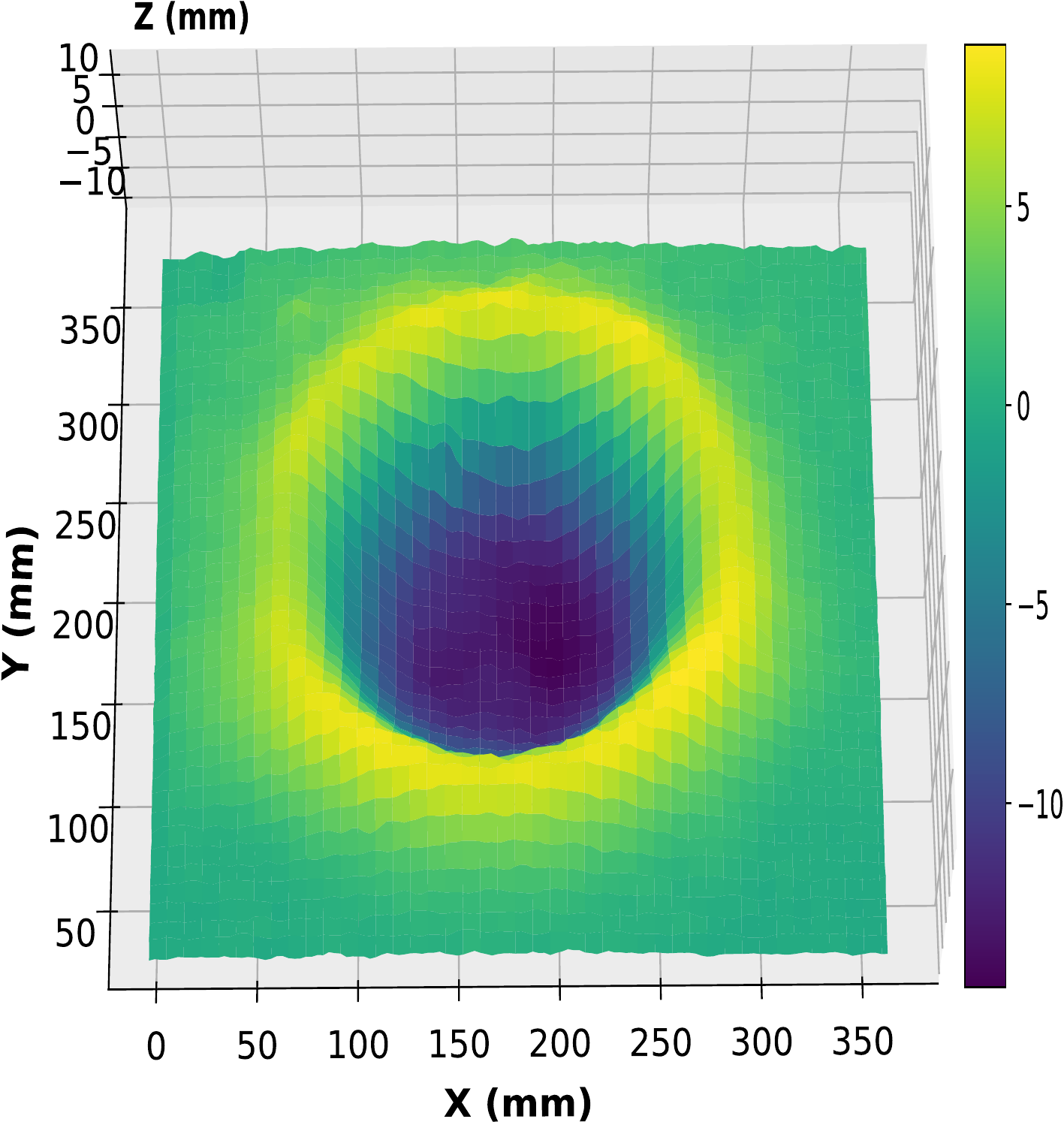}
\caption{}
\label{fig:5a}
\end{subfigure}
\begin{subfigure}[ht]{0.3\textwidth}
\centering
\includegraphics[width=\textwidth]{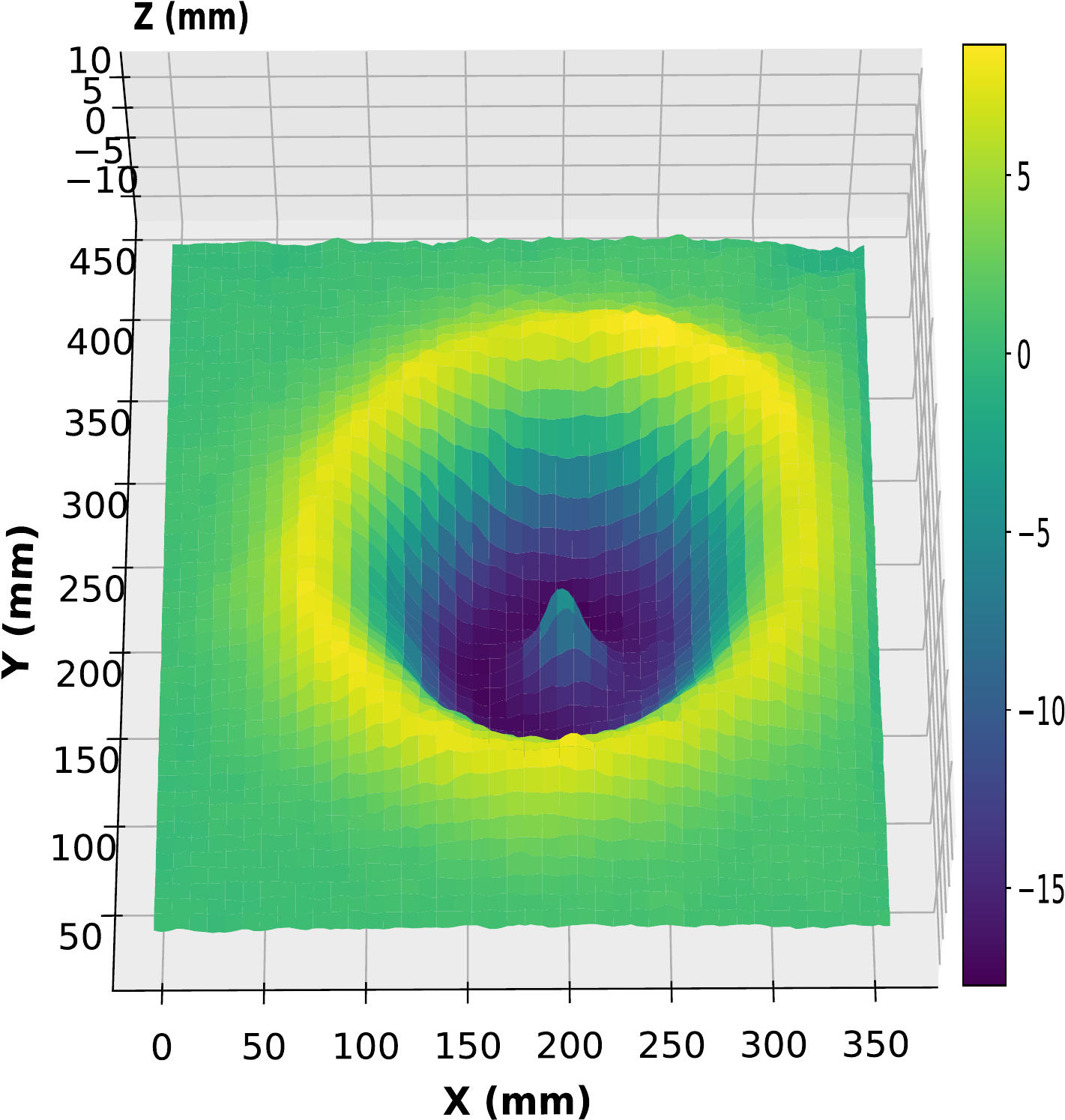}
\caption{}
\label{fig:5b}
\end{subfigure}
\begin{subfigure}[ht]{0.3\textwidth}
\centering
\includegraphics[width=\textwidth]{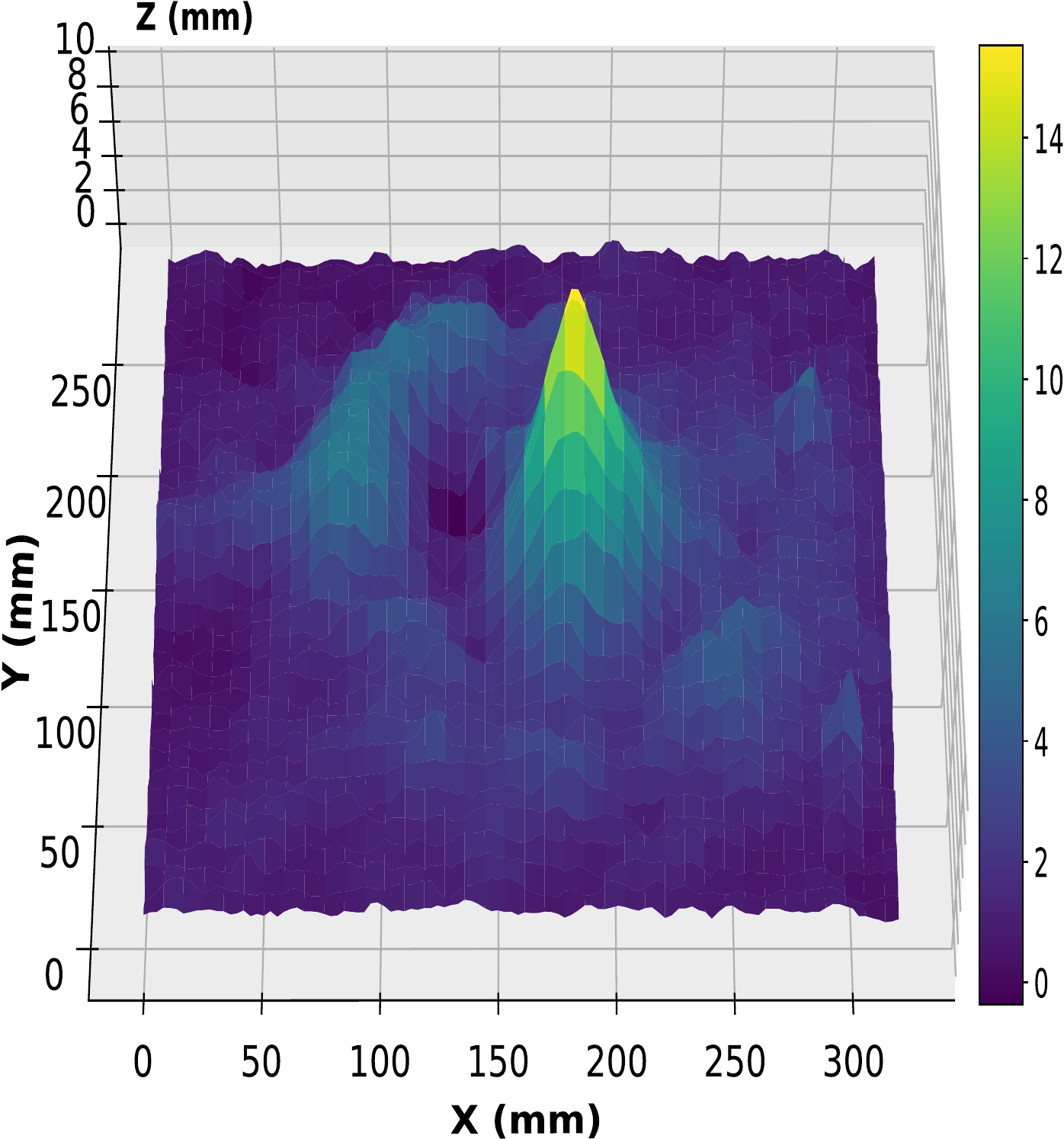}
\caption{}
\label{fig:5c}
\end{subfigure}		
\begin{subfigure}[ht]{0.3\textwidth}
\centering
\includegraphics[width=\textwidth]{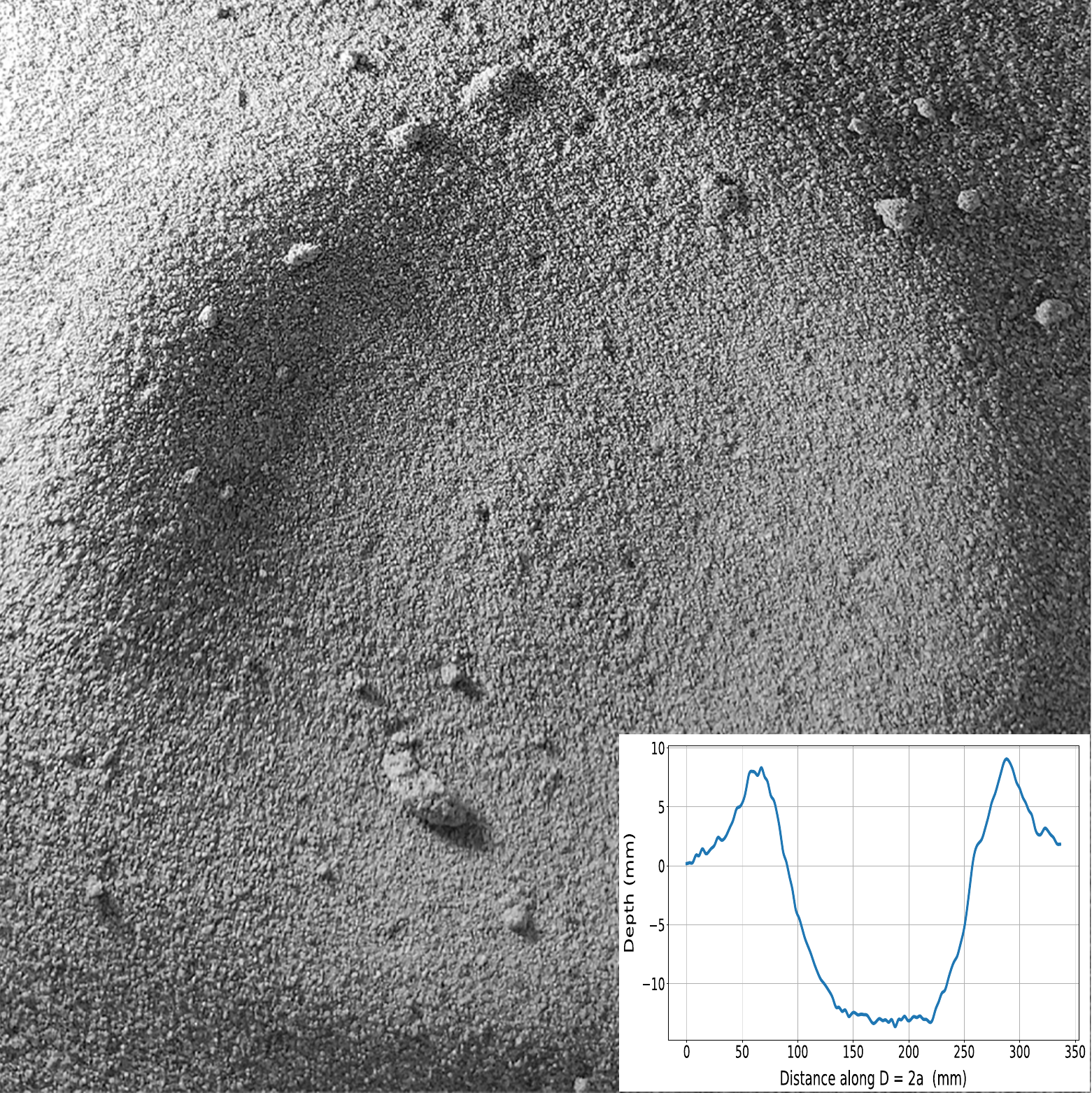}
\caption{}
\label{fig:5d}
\end{subfigure}
\begin{subfigure}[ht]{0.3\textwidth}
\centering
\includegraphics[width=\textwidth]{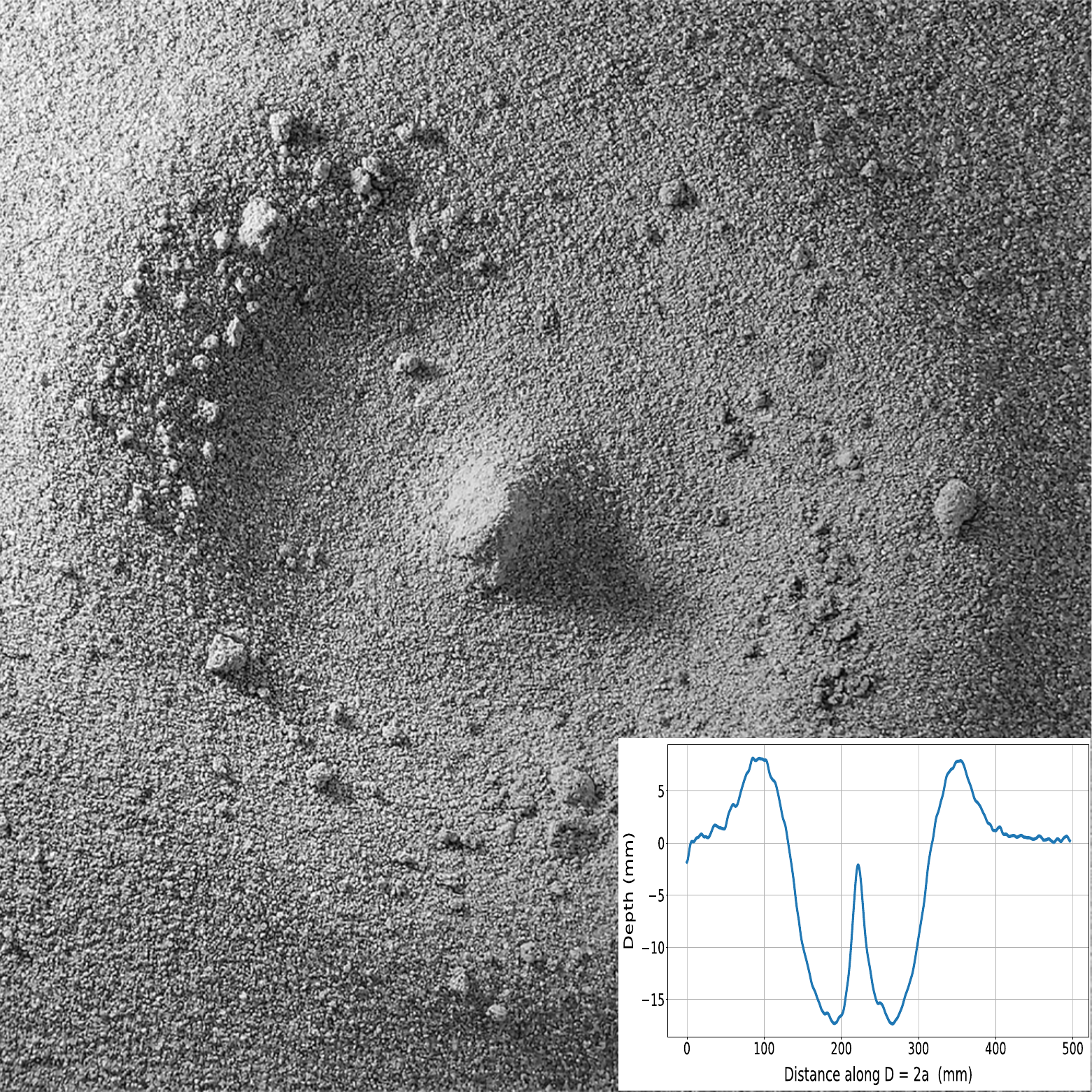}
\caption{}
\label{fig:5e}
\end{subfigure}
\begin{subfigure}[ht]{0.3\textwidth}
\centering
\includegraphics[width=\textwidth]{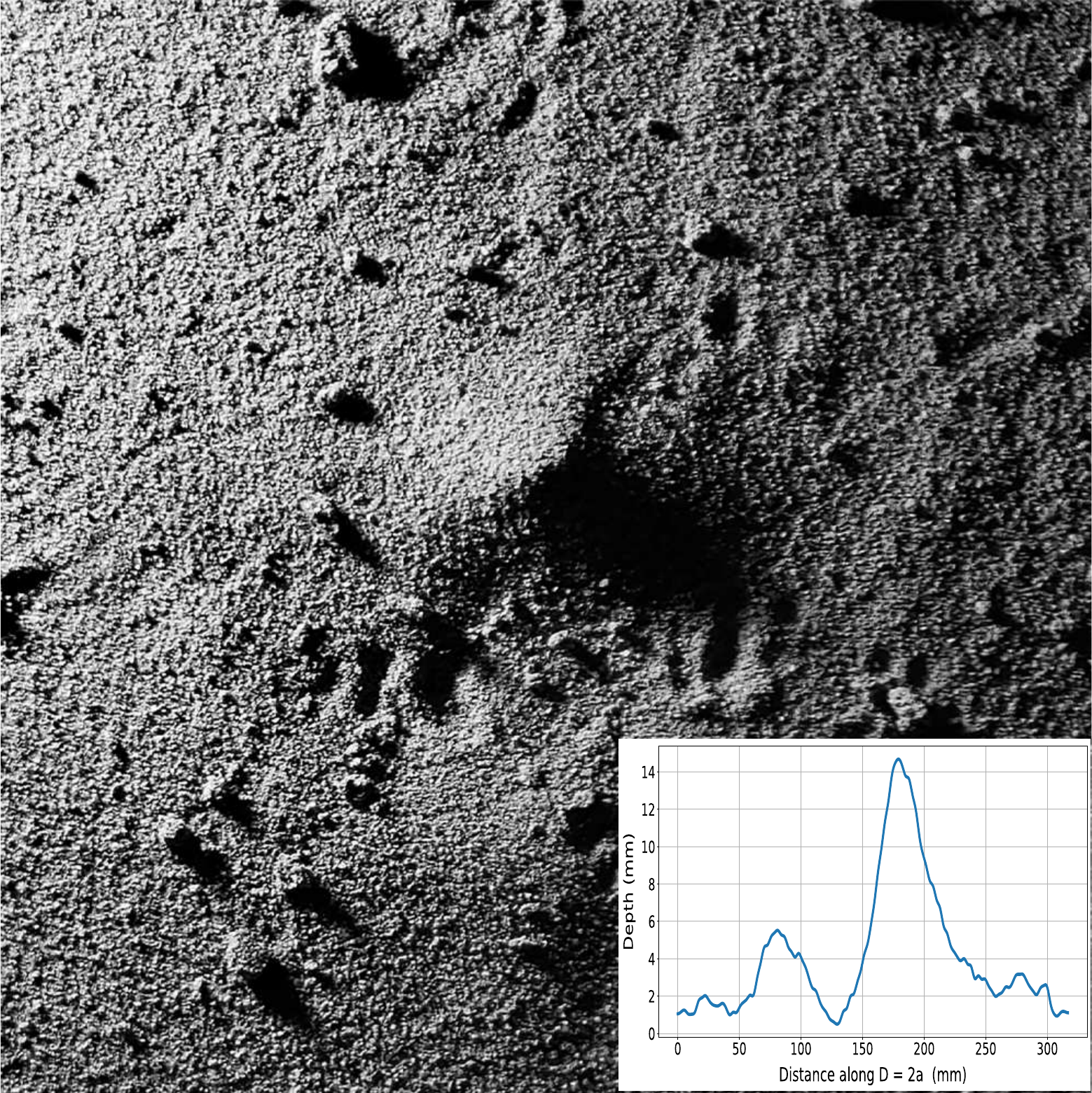}
\caption{}
\label{fig:5f}
\end{subfigure}
\begin{subfigure}[ht]{0.3\textwidth}
\centering
\includegraphics[width=\textwidth]{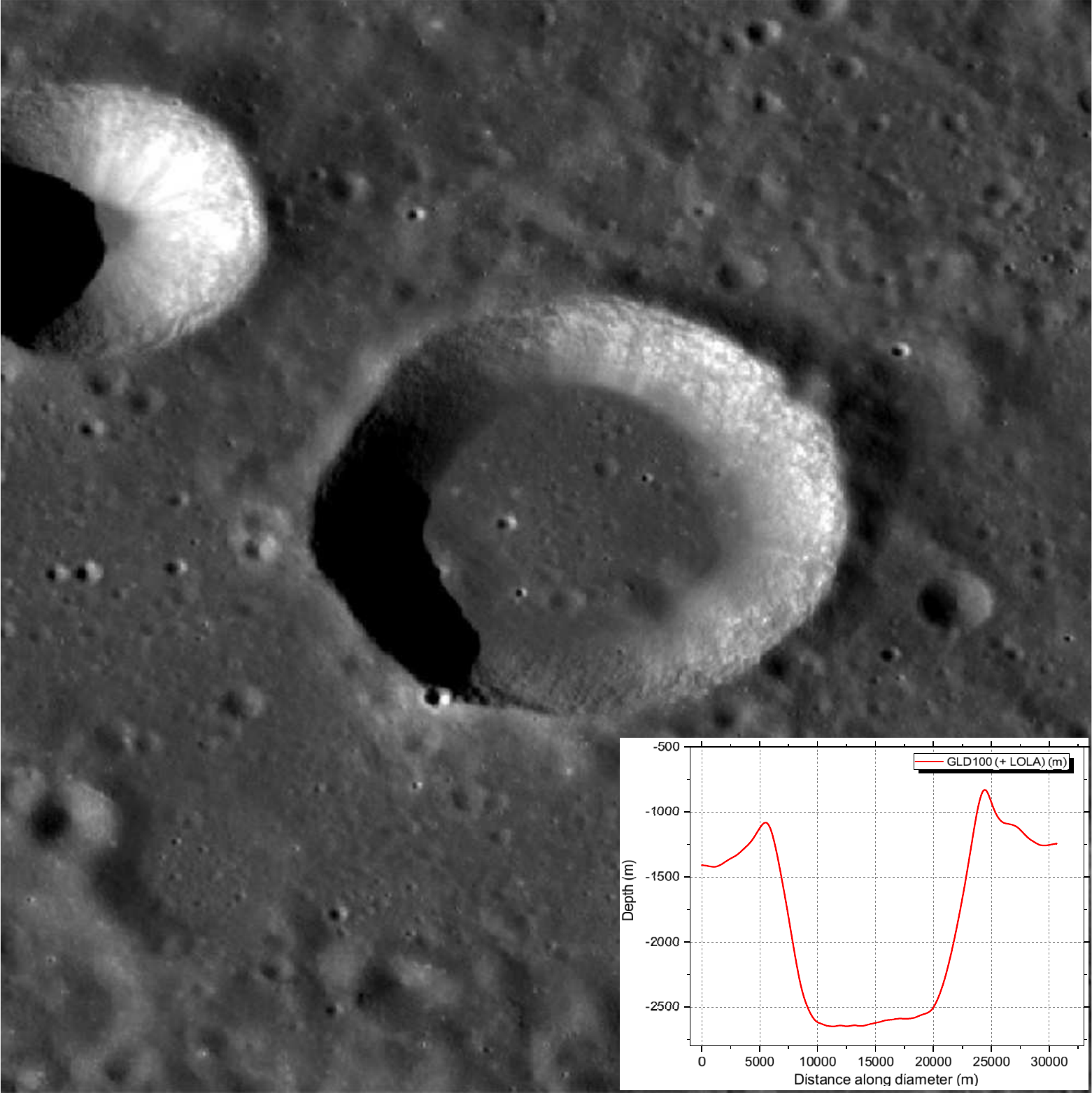}
\caption{}
\label{fig:5g}
\end{subfigure}
\begin{subfigure}[ht]{0.3\textwidth}
\centering
\includegraphics[width=\textwidth]{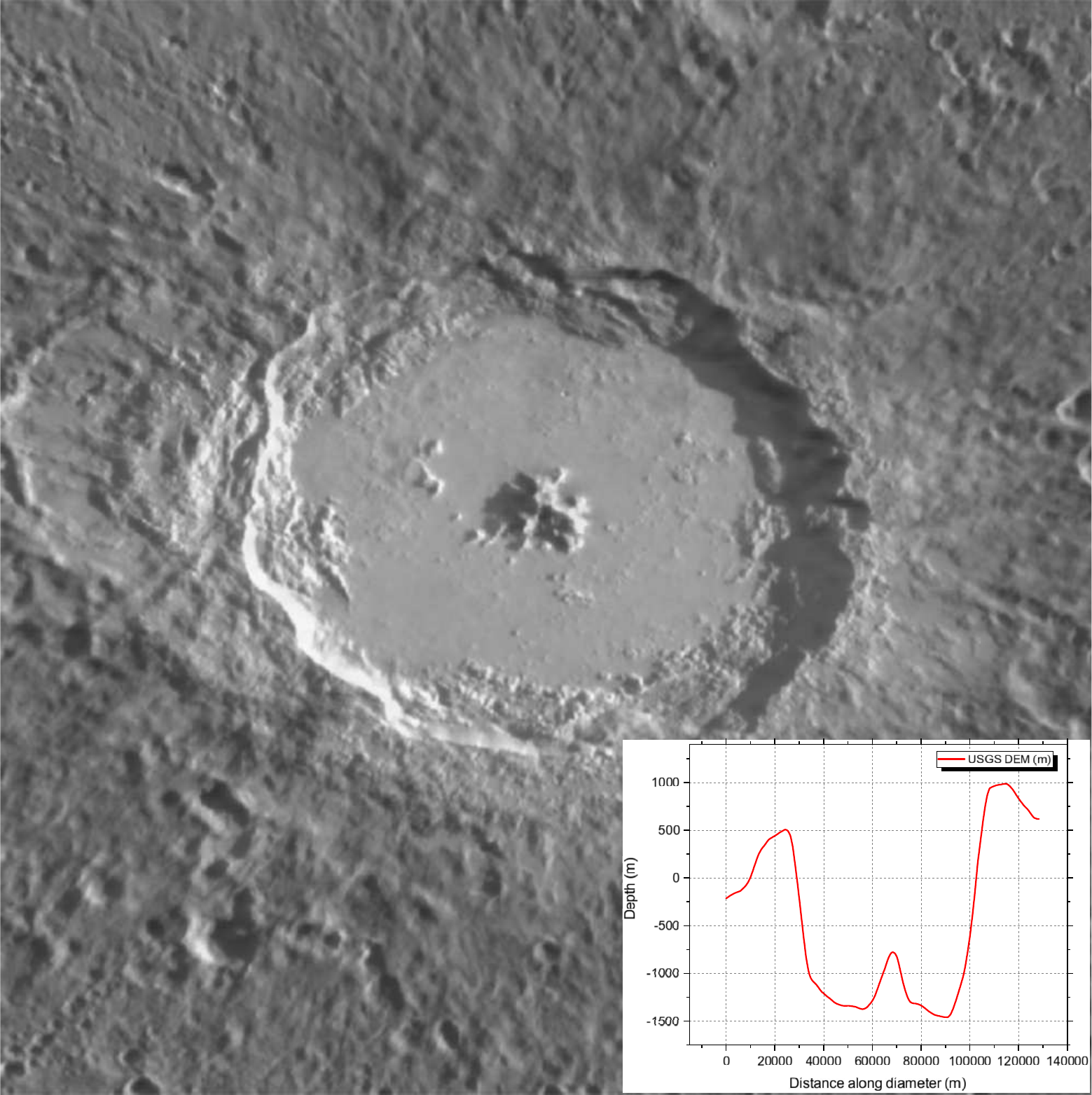}
\caption{}
\label{fig:5h}
\end{subfigure}
\begin{subfigure}[ht]{0.3\textwidth}
\centering
\includegraphics[width=\textwidth]{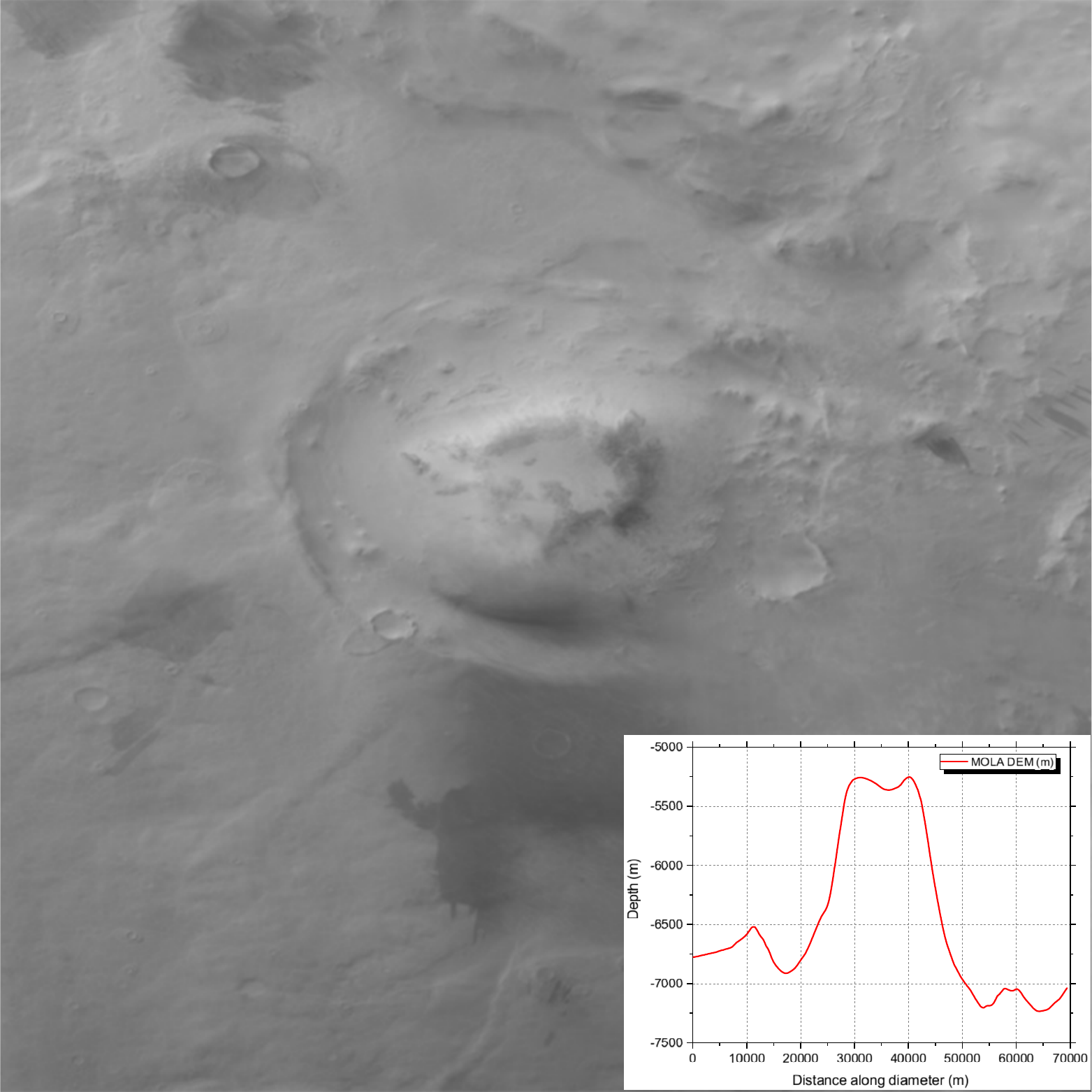}
\caption{}
\label{fig:5i}
\end{subfigure}
\caption{Three-dimensional visualization of the morphological surface of granular impact craters using \emph{craterslab} with Kinect depth data. Experimental and natural crater images are shown for comparison.
(a) A simple crater obtained at a height of $7.0$\,m above a loose packed granular bed with $V_{in}= 442966.88$\,mm$^3$. 
(b) A complex crater obtained at a height of $9$\,m above a loose packed granular bed with $V_{in} =  550837.34$\,mm$^3$. The morphology inside the crater changes compared to (a) due to a slight variation in potential energy. 
(c) Sand mound formed by the remnants of the projectile on the compacted impact surface at a height of $2$\,m with $V_{exc}= 191267.39$\,mm$^3$. 
  The experimental images (d), (e), and (f) correspond to the reconstructed three-dimensional models and serve as a visual comparison, showcasing the similarities and differences between the experimental craters and their reconstructions using ToF sensors. The insets correspond to the cross-sectional profile obtained by \emph{craterslab}. Similarly, images (g), (h), and (i) display natural craters alongside the insets of their cross-sectional profiles.
(g) Simple crater Bernoullie C on the Moon, inset extracted from the LROC platform using LOLA (Lunar Orbiter Laser Altimeter).
(h) Complex crater Debussy on Mercury, inset obtained from the USGS DEM (United States Geological Survey Digital Elevation Model).
(i) Mound formation on Mars without nomenclature, coordinates: Latitude: $-45.47963$, Longitude: $55.10807$, inset extracted from the MOLA DEM (Mars Orbiter Laser Altimeter Digital Elevation Model).
}
\label{fig:Figure5}
\end{figure*}

In order to validate both, the methodology for studying craters morphologies through ToF sensor and our software library for automating the process, we conducted a set of experiments at different launching heights on a compacted or loose packed sand bed as described in section \ref{sect:materiales}, producing a wide range of impact craters types or sand mounds. Figure \ref{fig:Figure5} shows the outcomes of three different type of morphologies produced experimentally. 
	
Figure \ref{fig:5a} shows the resulting data gathered and visualized using our software for the case of a simple crater, similarly, Figure \ref{fig:5b} and Figure \ref{fig:5c} resemble the data from a complex crater and a sand mound respectively. For all three cases, we included an image of the surface taken after the impact. Those images can be seen in Figures \ref{fig:5d}, \ref{fig:5e} and \ref{fig:5f} respectively. When comparing the images in the first and second rows of Figure \ref{fig:Figure5}, the remarkable similarities between the experimental craters and their three-dimensional visualizations by \emph{craterslab} are evident.

The images in the third row (Figures \ref{fig:5g}, \ref{fig:5h}, \ref{fig:5i}) depict natural craters on the Moon, Mercury, and Mars, respectively \citep{lroc_bernoullie,lroc_debussy,lroc_noname}. They were included to highlight the similarities found between our experimental craters and those in our solar system. The insets of these images represent the cross-sectional profiles obtained from the platforms provided by Applied Coherent Technology (ACT) Corporation. Upon comparing the profiles of the images in the second row, obtained by \emph{craterslab}, with those in the third row of Figure \ref{fig:Figure5}, the striking similarity between granular analog craters and natural craters is remarkable. 

\begin{figure*}[ht]
\centering
\begin{subfigure}[ht]{0.24\textwidth}
\centering
\includegraphics[width=\textwidth]{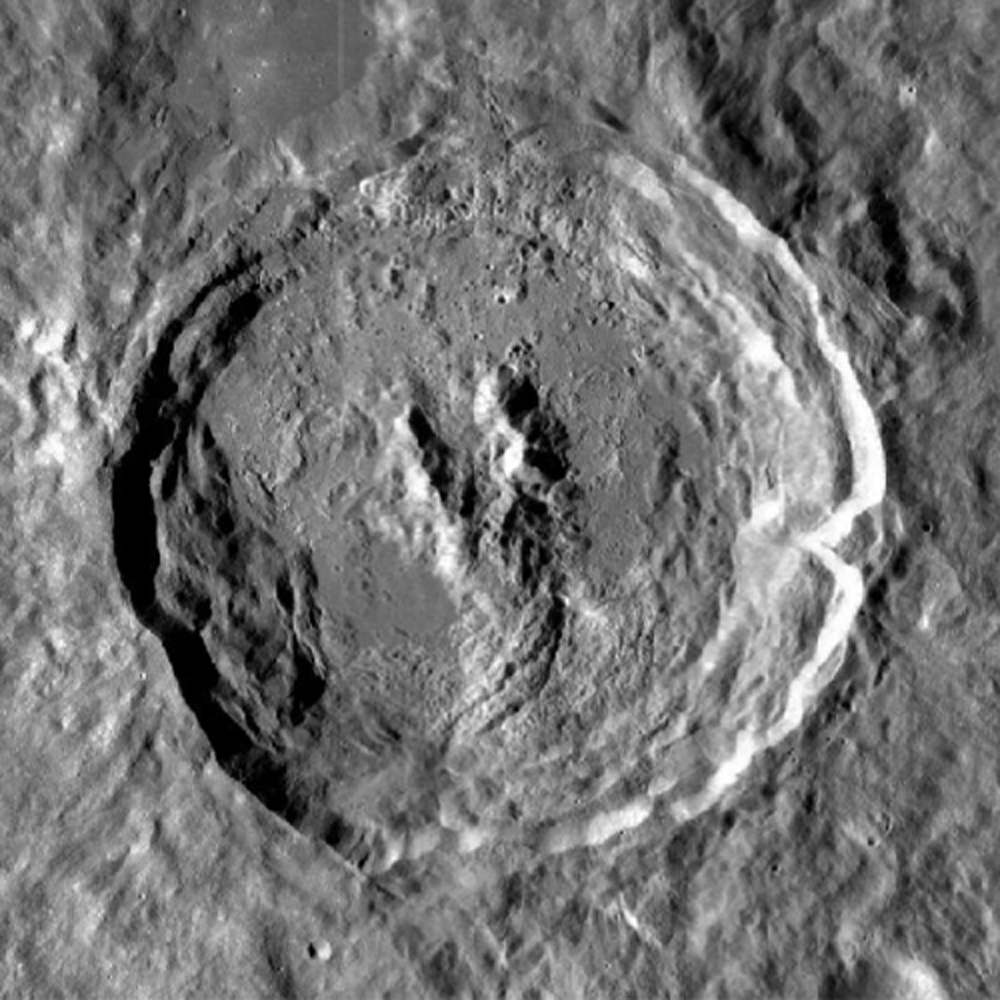}
\caption{}
\label{fig:6a}
\end{subfigure}
\hspace{1cm}
\begin{subfigure}[ht]{0.26\textwidth}
\centering
\includegraphics[width=\textwidth]{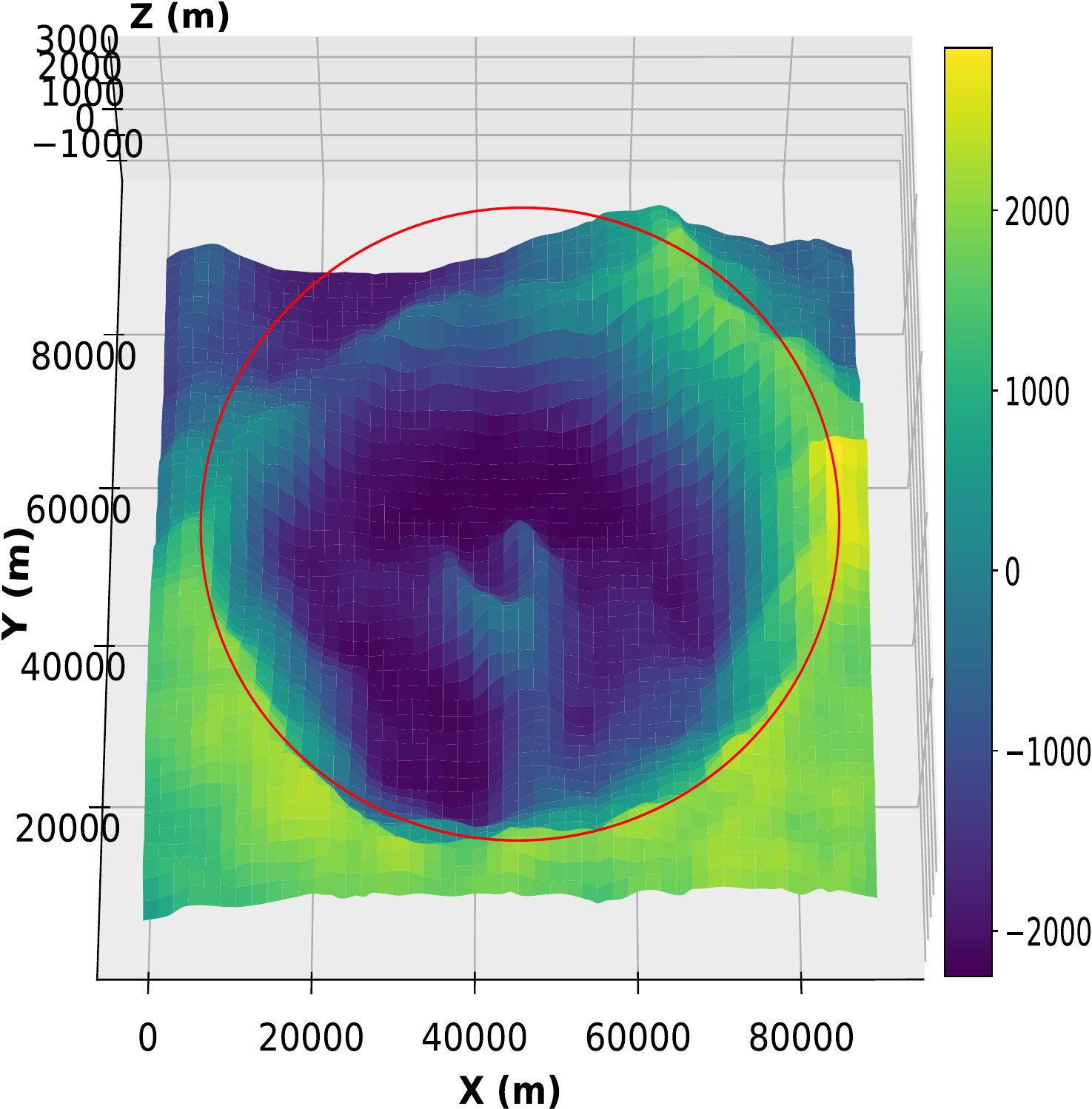}
\caption{}
\label{fig:6b}
\end{subfigure}
\hspace{1cm}
\begin{subfigure}[ht]{0.35\textwidth}
\centering
\includegraphics[width=\textwidth]{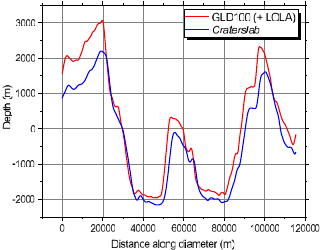}
\caption{}
\label{fig:6c}
\end{subfigure}	
\caption{Depth map analysis of the King crater using \emph{craterslab}.
(a) King Crater, with a diameter of $77$\,km and a depth of $5$\,km, is one of the youngest craters on the far side of the Moon and serves as an excellent example of a Copernican-aged complex impact crater. 
(b) Three-dimensional projection of King Crater. The volume of its cavity is $V_{in} = 5392.65$\,km$^3$. 
(c) Comparison of cross-sectional profiles of King Crater obtained from the LROC platform using LOLA (Lunar Orbiter Laser Altimeter) and \emph{craterslab}
}
\label{fig:Figure6}
\end{figure*}

In order to expand the evaluation, we proceeded to use our software for analyzing the depth map of the King crater, a well known lunar crater. The results are presented in Figure \ref{fig:Figure6}, where a three-dimensional visualization of the crater surface with the fitted ellipse on the crater rims (Figure \ref{fig:6b}) is displayed. In addition to reproducing natural craters in three dimensions and enabling visual analysis, \emph{craterslab} is also capable of extracting the main observables that allow for the analysis of their morphological characteristics. 

For the King crater (Figure \ref{fig:6a}) \citep{lroc_king}, our software provides results that can be directly compared with those from the LROC platform, such as cross-sectional profile and interior slopes. However, \emph{craterslab} can obtain and analyze additional observables from natural craters, for example the cavity volume ($V_{in}$).

Additionally, a comparison is shown between the profile view of the crater over the ellipse's largest radii obtained by \emph{craterslab} and the profile view by LROC, Figure \ref{fig:6c}.	The profile for the King crater obtained by the software is similar to the one obtained by the LROC tool. The slight differences can be attributed to the manual selection of the profile with the LROC tool, which does not allow for obtaining the largest profile from the crater automatically.
 
Once the 3D data from the experimental or natural craters is obtained, the software can compute the main observables automatically, eliminating the need for manual calculations or laborious image analysis procedures. This automation not only saves time but also ensures a more reliable and consistent analysis, leading to a deeper understanding of the crater morphology and how is correlated with its associated launching parameters.

Next, we obtain and characterize the morphological variations of the impact surface using our library and a KinectToF as the depth sensor. All craters produced by the collision events were  characterized by both techniques: profilometry and software-aided depth map analysis. The morphological characteristics of craters were measured manually from upper view pictures for profilometry, but both manually and automatically from the depth maps provided by the Kinect sensor, for comparison purposes.

\begin{figure*}[ht]
\centering
\begin{subfigure}[ht]{0.32\textwidth}
\centering
\includegraphics[width=\textwidth]{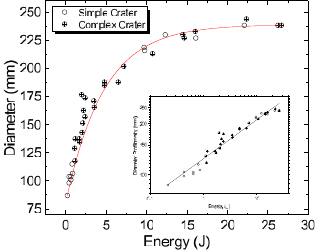}
\caption{}
\label{fig:7a}
\end{subfigure}		
\begin{subfigure}[ht]{0.32\textwidth}
\centering
\includegraphics[width=\textwidth]{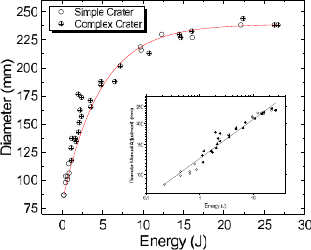}
\caption{}
\label{fig:7b}
\end{subfigure}		
\begin{subfigure}[ht]{0.32\textwidth}
\centering
\includegraphics[width=\textwidth]{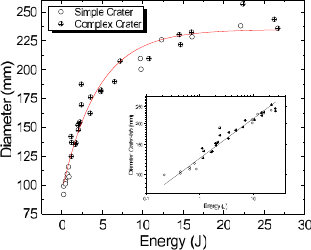}
\caption{}
\label{fig:7c}
\end{subfigure}
\caption{Diameter vs. Potential Energy for Impacts on a Loose Packed Granular Bed. The insets display logarithmic scale plots accompanied by linear fits. All linear fits cases exhibit a slope of $0.23$. This preliminary result is close to the exponent found in the relationship $D \propto E^{1/4}$ for natural craters in our solar system.
(a) Diameters obtained using the profilometry method. 
(b) Diameters estimated manually using Kinect data. 
(c) Diameters computed automatically using \emph{craterslab}.}
\label{fig:Figure7}
\end{figure*}

For the Kinect measurements in the plane $X,Y$, we first determined the craters diameter manually, mimicking the processing conducted with the profilometry, and then we automatically fitted a rotated ellipse using our software in order to compare both methods for measuring the diameter of the crater. Results are shown in Figure \ref{fig:7b} and Figure \ref{fig:7c} respectively. It is notorious that both methods are equivalent for diameter purposes, at least for the eccentricity values of these normal incidence impact craters.

Comparing the results from the profilometry technique (see Figure \ref{fig:7a}) with the manual processing of the depth map (see Figure \ref{fig:7b}), a standard deviation of $0.028$\,mm is obtained for distance values. This indicates that, under our working conditions, the resolution of the Kinect camera is equivalent to the profilometry method in the $X,Y$ plane.
	
Consequently, the differences in diameter size obtained from the software, manual estimation from the depth map, and profilometry are nearly indistinguishable, as depicted in Figure \ref{fig:Figure7}. 

\begin{figure}[ht]
\centering
\begin{subfigure}[ht]{0.23\textwidth}
\centering
\includegraphics[width=\textwidth]{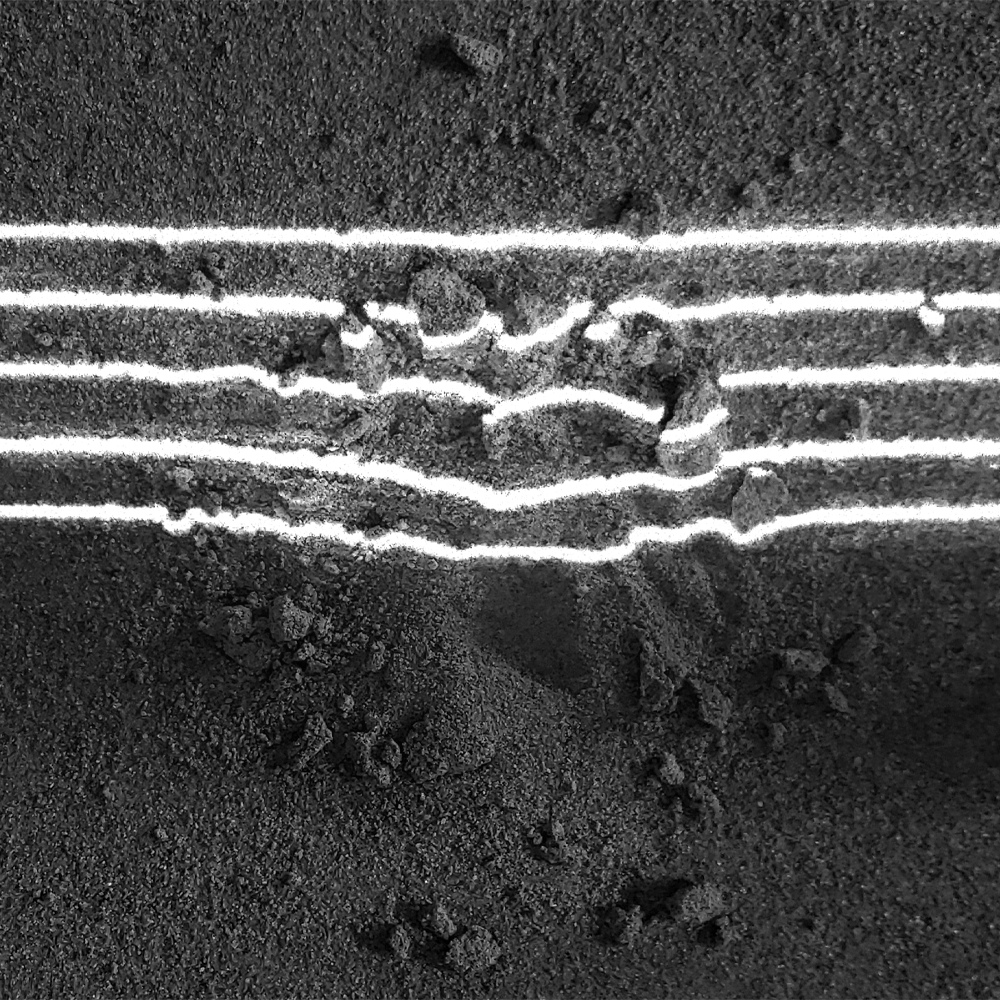}
\caption{}
\label{fig:8a}
\end{subfigure}
\begin{subfigure}[ht]{0.23\textwidth}
\centering
\includegraphics[width=\textwidth]{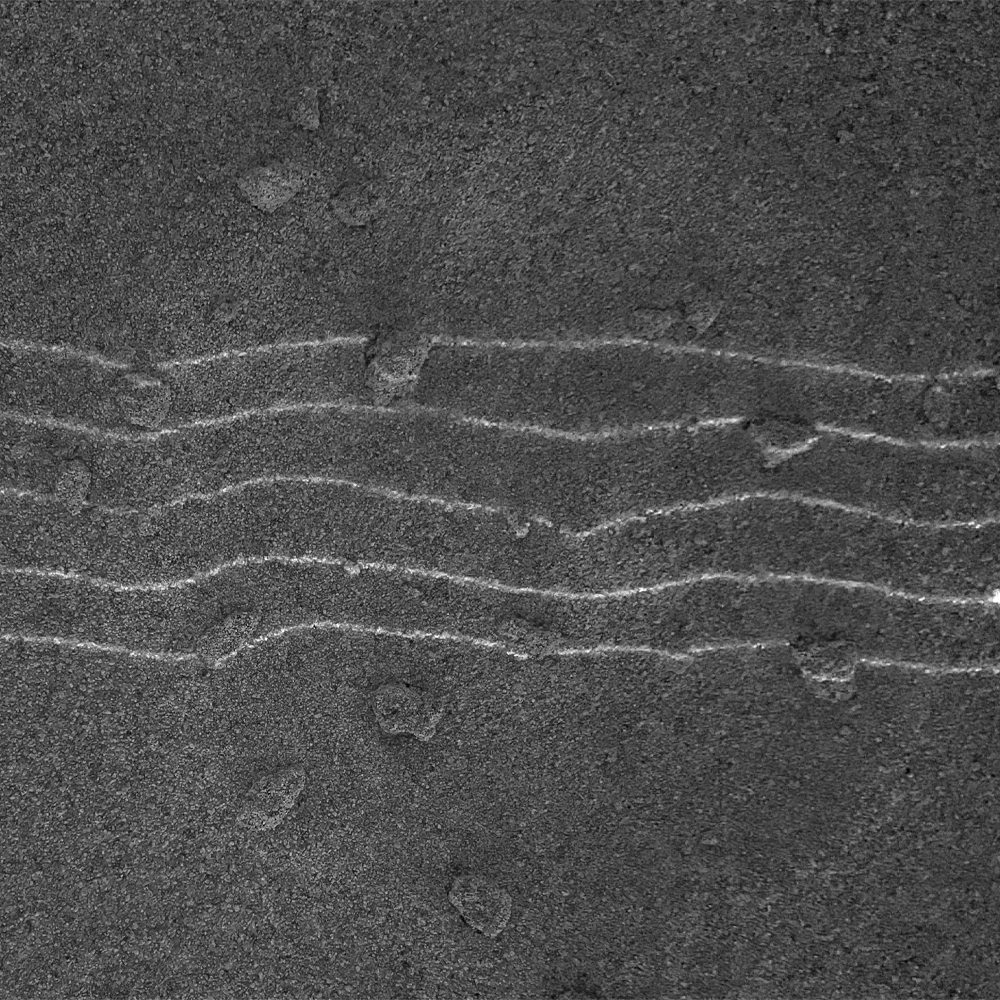}
\caption{}
\label{fig:8b}
\end{subfigure}
\caption{Lines of the laser light beam inside the crater depending on the lighting conditions. 
(a) thickness $6.96$\,mm.
(b) thickness $3.36$\,mm.}
\label{fig:Figure8}
\end{figure}

In morphological characterizations involving the $Z$ plane, both techniques are equivalent for obtaining depth data but profilometry exhibits a higher margin of error compared to KinectToF. The increased errors in profilometry occur within the interior of the fitted ellipse. This is attributed to the granular nature of the surface and the lighting conditions on the impact surface, which cause the thickness of the laser lines to increase within the crater. This introduces greater uncertainty in the measurement of depth values, as depicted in Figure \ref{fig:Figure8}. The average thickness in error  of the laser lines within the crater is $5.09$\,mm.

In contrast, Kinect depth data exhibits an offset of $\pm1$\,mm in our measurements. This offset represents the correction applied to align the depth measurements with the true surface positions, compensating for any systematic errors introduced by the sensor or experimental setup. 
	
As a result, Kinect provides higher precision in the three-dimensional reconstruction of granular impact surfaces.

\section{Conclusions}	

We propose a methodology for studying impact craters based on Time-of-Flight sensors. We validate our approach by comparing it with the established technique which relies on profilometry. 
Surface topographic data are gathered using a KinectToF camera for different impacting energies and compaction of the target terrain, producing a variety of crater shapes whose main morphological features are recognized and automatically measured, including; shape, depth, local slope, excess and excavated volumes as well as central peak volume and height. KinectToF exhibits high precision for the task, outperforming laser profilometry.

A software for automating the process is released as part of this work. It is designed for data acquisition and analysis of granular impact craters and natural craters. Time spent on both acquisition and analysis of data is considerably minimized when compared with previous methods due to the usage of this software. 
	
LiDAR sensors, specifically the second generation of Microsoft Kinect, when combined with our software, are able to provide very accurate results from craters morphology. As a consequence, previously used geometric approximations, such as cavity's volume, can now be calculated numerically with greater precision. The automatic computation of other observables performed by the software, such as excess volume, excavated volume and inner slope of the crater, may be pivotal for advancing various research topics in field. For instance, determining projectile's penetration angle in craters formed by oblique impacts and analyzing the amount of material remaining as a deposit or ejected matter after impacts.

\textbf{Code availability section}

Name of the library: craterslab

Contact: gustavo.vieralopez@gssi.it

Hardware requirements: Any system compatible with Windows, Linux or MacOS

Program language: Python 3.10+
 
Software required: Python

Program size: Scripts size -- 34.2 kB, pretrained model -- 26.3 MB

The source code is available for downloading at the link:
https://github.com/gvieralopez/craterslab

\textbf{Data availability}

The data used for this work is available for downloading at the link:
https://github.com/gvieralopez/craters-data

\printcredits
	
\bibliographystyle{cas-model2-names}
\bibliography{cas-refs}

\end{document}